\documentstyle[12pt,epsf]{article}
\newcommand\pubnumber{SLAC-PUB-6938\\ CERN-TH/95-159\\ TAUP-2269-95}
\newcommand\pubdate{July, 1995}
\newcommand\pubtype{T}

\def\Title#1{\begin{center} {\Large #1 } \end{center}}
\def\Author#1{\begin{center}{ \sc #1} \end{center}}
\def\Address#1{\begin{center}{ \it #1} \end{center}}
\def\andauth{\begin{center}{and} \end{center}}
\def\submit#1{\begin{center}Submitted to {\sl #1} \end{center}}
\def\doeack{\footnote{Work supported by the Department of Energy,
                     contract DE--AC03--76SF00515.}}
\def\SLAC{Stanford Linear Accelerator Center\\
    Stanford University, Stanford, California 94309 USA}
\newcommand{\journal}[4]{{\sl #1}\ {\bf #2}, #3\ (#4)}
\newcommand\pubblock{\rightline{\begin{tabular}{l} \pubnumber\\
         \pubdate \\ (\pubtype) \end{tabular}}}
\newenvironment{Abstract}{\begin{quotation} \begin{center}
                       ABSTRACT
     \end{center}\bigskip  }{\end{quotation}}
\def\beq{\begin{equation}}
\def\eeq#1{\label{#1}\end{equation}}
\def\eeqn{\end{equation}}
\def\beqa{\begin{eqnarray}}
\def\eeqa#1{\label{#1}\end{eqnarray}}
\def\eeqan{\end{eqnarray}}
\def\CR{\nonumber \\ }
\def\leqn#1{(\ref{#1})}
\def\Acknowledgements{\bigskip  \bigskip {\begin{center} \begin{large}
             \bf ACKNOWLEDGEMENTS \end{large}\end{center}}}
\def\bar#1{\overline{#1}}
\def\Dslash{\not{\hbox{\kern-4pt $D$}}}
\def\notR{\not{\hbox{\kern-4pt $R$}}}
\def\Nf{{N_f}}
\def\Nc{{N_c}}
\def\Qb{{\bar{Q}}}
\def\bnf{{\bar{N_f}}}
\def\tNc{{\tilde N_c}}
\def\Bb{{\bar{B}}}
\def\bq{{\bar{q}}}
\def\dT{{T^{\dagger}}}
\def\dq{{q^{\dagger}}}
\def\dbq{{{\bar{q}}^{\dagger}}}
\def\half{{1\over 2}}

\def\D{{\cal D}}
\def\T{{\cal T}}
\def\L{{\cal L}}
\def\M{{\cal M}}
\def\mQ{{m_Q^2}}
\def\mg{{m_g}}

\def\Mg{{M_g}}
\def\tr{{\rm tr}}
\def\del{\partial}
\def\One{{\bf 1}}
\def\VEV#1{\left\langle{ #1} \right\rangle}
\def\ra{\rightarrow}
\def\plp{{\bar{\psi}}^i_a \lambda^a_b \psi^b_j}
\def\etal{{\it et al.}}
\def\hc{{\rm h.c.}}
\def\coback{\footnote{Work supported in part by the Israel Academy of Science}}
\def\ofack{\footnote{Work supported in part by the Clore Scholars Programme}}
\def\telack{\footnote{Work supported in part by the US-Israel Binational
Science Foundation and by GIF -- the German-Israeli Foundation for
Scientific Research}}
\def\shimack{\footnote{On leave of absence from the School of Physics,
Raymond and Beverly Sackler Faculty of Exact Sciences, Tel-Aviv University}}
\def\telaviv{School of Physics and Astronomy\\ Beverly and Raymond Sackler
Faculty of Exact Sciences\\ Tel Aviv University,
    Ramat Aviv, Tel Aviv 69978, Israel}
\def\CERN{CERN, Geneva, Switzerland}
\begin{document}
\begin{titlepage}
\pubblock

\vfill
\Title{Exotic Non-Supersymmetric Gauge Dynamics from \\ Supersymmetric
            QCD}
\vfill
\Author{Ofer Aharony\ofack$^,$\coback$^,$\telack, Jacob Sonnenschein$^{2,3}$}
\Address{\telaviv}
\medskip
\Author{Michael E. Peskin\doeack}
\Address{\SLAC}
\medskip
\andauth
\medskip
\Author{Shimon Yankielowicz$^{3,}$\shimack}
\Address{\CERN}

\vfill
\begin{Abstract}
We extend Seiberg's qualitative picture of the behavior of supersymmetric
QCD to nonsupersymmetric models by
adding soft supersymmetry breaking terms.  In this
way, we recover the standard vacuum of QCD with $N_f$ flavors
and $N_c$ colors when $N_f < N_c$.  However, for $N_f \geq N_c$, we
find new exotic states---new vacua with spontaneously broken baryon number
for $N_f = N_c$, and a vacuum state with unbroken chiral symmetry for
$N_f > N_c$.  These exotic vacua contain massless composite fermions
and, in some cases, dynamically generated gauge bosons. In particular
Seiberg's electric-magnetic duality seems to persist also in the
 presence of (small) soft
supersymmetry breaking. We argue that certain, specially tailored, lattice
simulations may be able to detect the novel phenomena. Most of the exotic
behavior does not survive the decoupling limit of large SUSY breaking
parameters.
\end{Abstract}
\medskip
\submit{Physical Review {\bf D}}

\vfill
\end{titlepage}
\def\thefootnote{\fnsymbol{footnote}}
\setcounter{footnote}{0}
\section{Introduction}

When we ask how a gauge theory behaves at strong coupling, we want first
of all to understand how the chiral symmetry of this theory is realized.
In the familiar strong interactions, we know from experiment that the
approximate chiral $SU(3)\times SU(3)$ symmetry is
spontaneously  broken to a vector
$SU(3)$ symmetry.  This chiral symmetry breaking allows the quarks to
obtain dynamical masses and so  justifies the quark model
of hadrons.  For a long time, physicists have wondered whether this
same qualitative behavior should be found in any asymptotically free
gauge theory.  In a Yang-Mills theory in which the chiral symmetries
are not spontaneously broken, these unbroken symmetries can protect
composite fermions from obtaining masses \cite{tHooft}, leading to a
completely new dynamical picture.

In the early 1980's, this same question, which had not been resolved in
the case of ordinary Yang-Mills theory, was studied in the supersymmetric
extension of Yang-Mills theory.  In stages, the qualitative behavior was
worked out for supersymmetric pure Yang-Mills theory \cite{VY} and for
supersymmetric Yang-Mills theory with a small number of quark flavors
\cite{TVY,ADSlett,ADS}.  Recently, Seiberg has returned to this question
and, in a remarkable set of papers \cite{NatiMod,NatiDual}, has given a
coherent picture of the qualitative behavior of supersymmetric
QCD (SQCD)   for all
numbers of flavors.  Seiberg has emphasized that his solution includes
dynamical
features that are quite exotic, including vacuum states with baryon number
violation and massless composite fermions, and he has speculated that these
features can potentially also appear in nonsupersymmetric models.

In this paper, we will investigate the extension of Seiberg's vacuum states
to nonsupersymmetric models.  To do this, we will study how these vacua
are perturbed by the addition of soft
 supersymmetry breaking terms to the Lagrangian.  This
method is quantitative only when the soft supersymmetry breaking masses
are much smaller than the strong-coupling scale $\Lambda$ of the Yang-Mills
theory.  Despite this limitation, we will show that many of the exotic
features found by Seiberg, notably chiral symmetry realizations
and duality, do survive in softly broken nonsupersymmetric theories. We
will suggest the way in which the supersymmetric limit connects to
ordinary Yang-Mills theories of quarks alone.

Soft breaking of supersymmetric Yang-Mills theory was studied previously, with
a very different motivation, by Masiero and Veneziano \cite{MV,MPRV}.  We will
follow some of the route uncovered in their papers, but the recent
improved understanding of supersymmetric Yang-Mills theory will allow us to
obtain a more complete picture.

In addition to the intrinsic interest of exploring nonsupersymmetric
extensions of Seiberg's mechanism, this investigation has a broader
significance.  Today, the most important tool for investigating strong-coupling
gauge theories is numerical simulation on the lattice.  Up to
now, lattice gauge theory simulations have found evidence only for the
conventional pattern of chiral symmetry breaking.  However, it is by no
means clear that the current simulations have exhausted the possibilities
to be discovered.  Seiberg's work suggests that
lattice gauge theorists should look harder, and in theories with colored
fundamental scalars as well as fermions.   If supersymmetry were
essential to Seiberg's vacuum states, these states would be very difficult
to reproduce in simulations, since, in general, there is no
known method of ensuring supersymmetry on the lattice.\footnote{See,
however, \cite{ES}, where certain $N=2$
supersymmetric lattice theories have been considered.}
Thus, lattice gauge theorists could reasonably expect success in
demonstrating the presence of  massless composite fermions and other exotic
 features only if these phenomena
exist in nonsupersymmetric models. Our analysis provides evidence that they
do, and it suggests the particular nonsupersymmetric models which are
the most promising for finding them.

In this paper, we consider $SU(N_c)$ Yang-Mills theories coupled to $N_f$
flavors of quarks and squarks.  In Section 2, we define our notation
and set up a general strategy for analyzing these models. In Section 3, we
consider the case  $N_f < N_c$.  For this case, we show that soft breaking
of supersymmetry leads to the conventional pattern of chiral symmetry
breaking, $SU(N_f)\times SU(N_f)$ spontaneously broken to the diagonal
$SU(N_f)$.  In Section 4, we consider the case $N_f = N_c$.  In this case,
we find that this conventional vacuum state still exists, but that a new
vacuum state also appears, with massless composite fermions and spontaneously
broken baryon number.

  In Section 5, we consider the case $N_f = (N_c +1)$.
In this case, we find that, for small soft supersymmetry breaking terms,
the chiral symmetry remains unbroken.  The vacuum state of this theory
contains massless composite fermions with quark and squark constituents; these
remain massless even when the squarks have nonzero mass, illustrating
a possibility for composite states first discussed by Preskill and
Weinberg \cite{Weinberg,DandP}.
In Section 6, we discuss the case $N_f \geq (N_c + 2)$.  Here the physics of
chiral symmetry breaking is quite similar to that found in the previous
situation.  Seiberg has argued that the supersymmetric limit of these
models also possesses a dynamically generated  gauge symmetry which,
in some circumstances, is weakly coupled. This gauge symmetry is often lost
in the nonsupersymmetric case, but we will give some specific models in
which it survives. In particular, it seems that the electric-magnetic
duality   which Seiberg claimed for this region
 persists  in the presence of (small) soft
supersymmetry breaking.

  Most of our discussion will be carried out for the case $\Nc \geq 3$.
The case $\Nc = 2$ has a number of special
complications.  However, since
this is the case of most interest to people with computers of finite
capacity, we discuss this case specifically in Section 7.
Lattice simulations of gauge theories with scalar fields have a
practical difficulty that it may not be possible to reach the continuum
limit, due to the presence of a first order phase transition as a
function of the scalar field mass parameter.  In Section 8, we discuss
how this problem can arise from  perturbation of the  supersymmetric
Lagrangian, and how it can be avoided.

  In all, these models open a wide variety of new phenomena in
nonsupersymmetric models, raising many  possibilities for theoretical and
numerical investigation and for model-building.  They confirm Seiberg's
intuition that, while supersymmetry is useful for investigating the variety
of behaviors  possible in strongly coupled gauge theories, it is not a
necessary  condition for their realization.

\section{Notations and Strategy}

In this paper, we will be concerned with $SU(N_c)$ Yang-Mills
theories coupled to $N_f$ flavors of quarks.  We will be perturbing about the
supersymmetric limit of these theories.  In this limit, these theories contain
fundamental scalar (squark) fields and a fermion (gluino) in the adjoint
representation of the gauge group, in addition to the standard content of
Yang-Mills theories with fermions.

\subsection{Fields and Symmetries}

 The quarks and squarks can be grouped into
chiral superfields in the $N_c$ and $\bar{N_c}$ representations of $SU(N_c)$.
We will refer to these superfields as
\beq
    Q^i_a \ , \qquad  \Qb^a_i  \ ,
\eeq{squarks}
where $i= 1, \ldots, \Nf$ is a flavor index
and $a = 1, \ldots, \Nc$ is a color index.  When we wish to refer to the
individual components of the superfield, we will denote the scalars  by
$Q$, $\Qb$ and the fermions by $\psi_Q$, $\psi_{\Qb}$.
The Hermitian conjugate superfields will be denoted $Q^\dagger, \Qb^\dagger$.
Note that while
$\psi_Q$ is a left-handed quark, $\psi_{\Qb}$ is a left-handed antiquark; the
right-handed quarks are components of $\Qb^\dagger$.
We will reserve the notation $q$, $\bar q$ to denote Seiberg's dual quark
superfields, which will appear in Section 6.  We will denote the gluino as
$\lambda^a{}_b$, a matrix in the
 adjoint representation of $SU(N_c)$.

When $\Nc = 2$, the representations $\Nc$ and $\bar{\Nc}$ become equivalent,
 and this introduces a number of complications.  From this introduction
 through Section 6, we will
restrict ourselves to $\Nc \geq 3$.  In Section 7, we will discuss the
generalization of our results to $\Nc = 2$.

In the classical SQCD theory
the quark superfields have no interactions beyond their
couplings to the gauge supermultiplet. In particular, we will assume that they
have zero mass.  This implies that the supersymmetric
 theory has a global symmetry
\beq
   SU(N_f)_L \times SU(N_f)_R \times U(1)_B \times U(1)_R ,
\eeq{globalsymm}
where $SU(\Nf)_L$ acts on the $Q^i$, $SU(\Nf)_R$ acts on the $\Qb_i$, and
$U(1)_B$ denotes baryon number.  We will refer to the vectorial flavor
group, the diagonal subgroup of the two $SU(\Nf)$'s, as $SU(\Nf)_V$.
The additional factor
$U(1)_R$ denotes the anomaly-free
combination of the axial $U(1)$ symmetry acting on the quarks and the
canonical $R$ symmetry which acts on all fermion fields. Under this
anomaly-free
symmetry, the squarks, quarks, and gluinos have the following charges:
\beq
Q, \Qb\ : \  {\Nf - \Nc\over \Nf}\qquad \psi_Q, \psi_\Qb\ :\ -{\Nc\over \Nf}
 \qquad \lambda\ :\  1 .
\eeq{Rcharges}
A superpotential $W$ should have $R$ charge 2.

\subsection{Effective Lagrangians}

The qualitative behavior of supersymmetric Yang-Mills theory is made most
clear by writing an effective Lagrangian in terms of gauge-invariant chiral
superfields.  As Seiberg especially has emphasized \cite{NatiMod}, this
Lagrangian is strongly constrained by the condition that its superpotential
must be a holomorphic function of these fields.  For small values of $\Nf$,
the only possible gauge-invariant chiral field built from the quark fields
 is the meson field
\beq
      T^i{}_j =  Q^i \cdot \Qb_j.
\eeq{mesons}
Beginning at $\Nf = \Nc$, there are also chiral superfields with the quantum
numbers of baryons.  Let
\beq
       \tNc = (\Nf - \Nc).
\eeq{Ntilde}
Then there is a baryon chiral superfield in the $\tNc$-index antisymmetric
tensor representation of $SU(\Nf)_L$,
\beq
     B_{i_1 \cdots i_\tNc} =   \epsilon^{a_1 \cdots a_\Nc}
\epsilon_{j_1 \cdots j_\Nc i_1 \cdots i_\tNc} Q^{j_1}_{a_1} \cdots
            Q^{j_\Nc}_{a_\Nc}  ,
\eeq{baryonop}
and, similarly, an antibaryon chiral superfield $\Bb^{i_1 \cdots i_\tNc}$
built from $\Nc$ powers of the field $\Qb$.

 Using the gauge supermultiplet,
it is possible to build another chiral superfield
\beq
     S = - \tr\bigl[ W^{\alpha } W_\alpha\bigr] = \tr[ \lambda \cdot
     \lambda]
                  + \cdots \ .
\eeq{Sdef}
The superfield $S$ has $R$ charge 2 and is neutral under the other
global symmetries.
In studies of the qualitative behavior of supersymmetric Yang-Mills theory,
the component fields of $S$ always acquire mass; these fields are
associated with the
massive hadrons of the pure glue sector of the theory.  However, the
dependence of the superpotential on $S$ is still fixed by symmetry arguments
\cite{VY,TVY}, and $S$ can be inserted or removed in an
unambiguous way  by Legendre transformations  \cite{ILS}. Though most
of our results can be derived without introducing $S$ into the
Lagrangian, it will be useful at some points in our analysis to write
effective Lagrangians that depend on $S$ as well as $T$.

\subsection{Soft Supersymmetry Breaking}

In addition to the superpotential, we will need to know the K\"ahler potential
 which determines the kinetic energy terms of the fields $T$, $B$,
and $\Bb$.  A simple hypothesis, introduced in the work of Masiero and
Veneziano \cite{MV,MPRV}, is that the K\"ahler potentials of the
gauge-invariant fields are canonical:
\beq
    K[T,B,\Bb] =   A_T \tr \bigl[ T^\dagger T \bigr] +
         A_B   \bigl( B^\dagger B + \Bb^\dagger \Bb \bigr) .
\eeq{cKahler}
Our main results will rely on weaker assumptions about the K\"ahler
potential, in particular, that it is nonsingular on the space of
supersymmetric vacuum states.  However, we will support our general
remarks by explicit calculations using this simple model.
We expect \leqn{cKahler} to be the correct form of the K\"ahler
potential near the origin of moduli space, in the cases for which the
mesons and baryons give an effective infrared  description of the theory.

We will also need to specify the terms by which we break supersymmetry.  In
this  paper, we will break supersymmetry by adding mass terms for the
squark fields and for the gaugino,
\beq
\Delta \L =   - \mQ \Bigl( \bigl| Q \bigr|^2 +  \bigl|\Qb \bigr|^2\Bigr)
                          + \bigl( \mg S  + {\rm h.c.}\bigr),
\eeq{softbr}
where, in \leqn{softbr}, $Q$, $\Qb$, and $S$ are the scalar component fields of
the superfields.  The scalar mass term is the unique soft supersymmetry
breaking term which does not break any of the global symmetries
\leqn{globalsymm} of the original model.  The gaugino mass term breaks only
the $U(1)_R$ symmetry, and thus breaks the global symmetry of the
supersymmetric model down to that of ordinary Yang-Mills theory with $\Nf$
massless
flavors.  Any other choice for the soft supersymmetry breaking terms would
induce further global symmetry breaking.  Because $S$ is a complex field, any
sign or phase inserted in front of the gluino mass term could be compensated
by a phase rotation of $S$ (or, more generally, by a $U(1)_R$
transformation).   We have chosen the phase of this term so that
the potential energy of the broken theory will be minimized when $S$ is real
and positive.

Actually, it is not clear whether the `correct' theory of broken supersymmetry
should or should not contain the gluino mass term.  If this term is included,
and then $\mQ$ and $\mg$ are taken to infinity, the theory reverts to the
standard Yang-Mills theory with $\Nf$ flavors.  If this term is omitted, and
then $\mQ$ is taken to infinity, the theory becomes a Yang-Mills theory
coupled to $\Nf$ flavors in the fundamental representation and one extra
flavor in the adjoint representation.  Both of these are theories whose
strong-coupling behavior might be of interest.  We will refer to the softly
broken theories without and with the $\mg$ term as the $R$ and $\notR$
theories, respectively.

Since we will be working in the language of the low-energy effective
Lagrangian, we must ask how the supersymmetry breaking term
\leqn{softbr} shows up in this Lagrangian.  To work this out, rewrite
\leqn{softbr} in the superfield form
\beq
    \Delta\L =  \int d^4\theta  M_Q \bigl( Q^\dagger e^V Q  +
					\Qb^\dagger e^{-V^T} \Qb \bigr)
                    + \int d^2\theta M_g S + {\rm h.c.} ,
\eeq{softbrsf}
where $M_Q$ is a vector superfield whose $D$ component equals $(-m_Q^2)$ and
$M_g$ is a chiral superfield whose $F$ component equals $m_g$.
It is straightforward to see that these superfields are
gauge-invariant and neutral under all of the global symmetries.

 The
effective Lagrangian description of $\Delta\L$ for $N_f \leq N_c+1$
is then given by writing
the most general Lagrangian built from $T$, $B$, $\Bb$ and a fixed
number of factors of $M_Q$ and $M_g$.  The supersymmetry breaking terms
have an ambiguity related to that of the K\"ahler potential, because many
possible invariant structures can be built from $T$,
$B$, and  $\Bb$.  In our explicit calculations,
we will assume that the coefficient of $M_Q$ is quadratic in these
fields; again, this assumption is precise near the origin of moduli
space.  Then the first order soft supersymmetry breaking terms in the
effective Lagrangian are
\beqa
 \Delta\L & = \int d^4 \theta  \Bigl(
    B_T M_Q \tr\bigl[ T^\dagger   T\bigr]  + B_B M_Q \big\{
     B^\dagger B + \Bb^\dagger \Bb\bigr\} + \bar{M_g} \M(T,B,\Bb) + \hc
      \Bigr) \CR
         & \quad + \int d^2 \theta  M_g \VEV{S} + \hc ,
\eeqa{SSoft}
where $\M(T,B,\Bb)$ is a function of the effective Lagrangian
superfields which is neutral under the global symmetries.
The quantity $\VEV{S}$ in \leqn{SSoft} should be
a combination of the effective Lagrangian
chiral superfields which has the quantum numbers of $S$. In general,
this condition restricts that function to be proportional to the
expectation value of $S$ as determined from the effective Lagrangian
of refs. \cite{VY,TVY} which includes $S$ as a basic field.  In some of
our examples, the symmetry of the vacuum will prohibit $S$ from
obtaining a vacuum expectation value; then the only effect of $M_g$
will be from the $D$-term in \leqn{SSoft}. The appearance of this unknown
$D$-term, however, will prevent us from making any quantitative predictions
after adding the gluino mass.

The squark mass terms in \leqn{SSoft} are not the most general terms that
can be written down. As in the K\"ahler potential \leqn{cKahler}, higher order
terms in the fields, suppressed by powers of $\Lambda$, may appear.
However, we expect \leqn{SSoft} to be approximately true near the origin
of moduli space $T=B={\bar B}=0$. Thus, whenever the vacuum which we
analyze will be near
the origin of moduli space (as will be the case for $N_f \geq N_c+1$), we
expect \leqn{cKahler} and \leqn{SSoft} to give a good quantitative
description of the theory. In other cases, notably for $N_f \leq N_c$
where some expectation values are expected to be of order $\Lambda$ or
higher, higher order terms cannot be neglected.
We expect that the qualitative
behavior which we will find when using
these simple terms will remain valid also in
the exact theory. However, we will not be able to trust the quantitative
results.

The ratio of coefficients $B_B/B_T$ will be important to our later
analysis, but this ratio
cannot be determined from the effective Lagrangian viewpoint.  At best,
we can argue naively that the coefficient of the mass term of a
composite field should be roughly proportional to the sum of the
coefficients of the mass terms of the constituents.  This would give
the relation
\beq
       B_B \approx {\Nc\over 2} B_T \ ,
\eeq{BBTratio}
which the reader might take as qualitative guidance.

         To avoid the proliferation of factors $\Lambda^\beta$,
where $\Lambda$ is the
nonperturbative scale of the strong interaction theory, we will generally
 choose units in which $\Lambda = 1$.  Then $\mQ$ and $\mg$ will be small
dimensionless numbers.  We
 emphasize again that our method makes
quantitative sense only for theories with weakly broken supersymmetry, that
is, only when $\mQ$ and $\mg$ are much less than
$\Lambda$ and will
not apply directly to models in which the squarks and
 gluinos are  completely decoupled.  However, in many of our examples, the
qualitative behavior we find in the region $\mQ \ll \Lambda$ will
suggest a smooth continuation to the decoupling limit $\mQ \gg \Lambda$.
In each case that we study, we will offer at least a plausible conjecture,
for both the $R$ and $\notR$ cases, of the  connection between
these two limits.

It is important for our analysis that the behavior of the theory is
non--singular when adding the squark and gluino masses, i.e. that no new
non--perturbative effects occur. In general it is not possible to prove
this in non--supersymmetric theories, but a proof of this is possible in
softly broken supersymmetric theories, when the soft breaking can be viewed
as spontaneous breaking of supersymmetry. For SQCD this was done by Evans
\etal\ \cite{Evans}, who showed how the squark and gluino mass terms may be
obtained by spontaneous supersymmetry breaking in a theory which includes
some additional chiral superfields. When obtaining the soft breaking terms
in this way, from a supersymmetric theory in which we have control over the
superpotential, we can show that the form of the SUSY breaking operators is
indeed as in equation \leqn{SSoft}. In fact, in \cite{Evans}, the squark
mass is derived from the K\"ahler term in the original SUSY theory, so that
our lack of control of this term in \leqn{SSoft} is related to our lack of
control over the K\"ahler term \leqn{cKahler}, and the two are expected to
behave in a similar fashion.

\section{$N_f < N_c$}

We begin with the simplest situation, $\Nf < \Nc$.  In this case, there are no
baryon operators; thus, in the supersymmetric limit, the only massless
particles are those created by the meson operator $T$.  In this section, we
will work out the vacuum and massless spectrum which result when this theory
is perturbed by the soft supersymmetry breaking terms \leqn{softbr}.

In this case, the effective theory of the supersymmetric limit is
described by the Affleck-Dine-Seiberg superpotential:
\beq
  \int d^2 \theta  W(T) = \int d^2\theta  {(\Nc-\Nf)\over (\det
T)^{1/(\Nc-\Nf)}},
\eeq{ADSpot}
where we have set $\Lambda = 1$ as described at the end of Section 2.
To begin, choose the
canonical K\"ahler potential \leqn{cKahler}.  We will
 comment on other choices of the K\"ahler potential below.
 Using \leqn{cKahler}, we find the
potential energy
\beq
    V(T) =  {1 \over A_T |\det T|^{ 2/(\Nc - \Nf)}}\tr\bigl[
          (T^{-1})^\dagger T^{-1}\bigr] .
\eeq{firstV}
Now add the soft supersymmetry breaking term \leqn{softbr}.
Again, we will begin with a simpler situation, choosing the $R$
case where $\mg = 0$.  The addition to the potential is
\beq
  \Delta V =  B_T \mQ \tr \bigl[ T^\dagger T\bigr] .
\eeq{delV}
To find the vacuum state, we must  minimize $V + \Delta V$.

\subsection{Location of the Vacuum State}

If we use the freedom of $SU(\Nf)\times SU(\Nf)$ to diagonalize $T$, this
potential can be written in terms of the  complex eigenvalues $t_i$ of
$T$, as
\beq
   V(T) = B_T \mQ \sum_i |t_i|^2 +
   {1\over A_T}{1\over |\prod t_i|^{2/(\Nc - \Nf)}}
         \cdot  \sum {1\over |t_i|^2}.
\eeq{secondV}
 The minimization
 equation is
\beq
   0 = B_T \mQ t_i  - {1\over A_T}{1\over \D^{2/(\Nc-\Nf)}}
 \Biggl[ {1\over t_i^* |t_i|^2} + {1\over (\Nc-\Nf) t_i^*}\T \Biggr],
\eeq{mineq}
where
\beq
   \D = |\prod_i t_i|\ ; \qquad \T = \sum_i {1\over |t_i|^2} .
\eeq{DTdefs}
Multiplying through by $t_i^*$, we find an equation of the form
\beq
  0 = B_T\mQ |t_i|^2 -  F(|t_i|^2, \D, \T) ,
\eeq{mineqtwo}
where, for fixed $\D$ and $\T$, the function $F$ decreases
monotonically as the first term increases monotonically from
0 to infinity.  This equation has a unique solution for $t_i$; thus,
all of the $t_i$ are equal at the minimum of the potential, up to
phases removable by global symmetry transformations.

\begin{figure}
\begin{center}
\leavevmode
{\epsfxsize=5.00truein \epsfbox{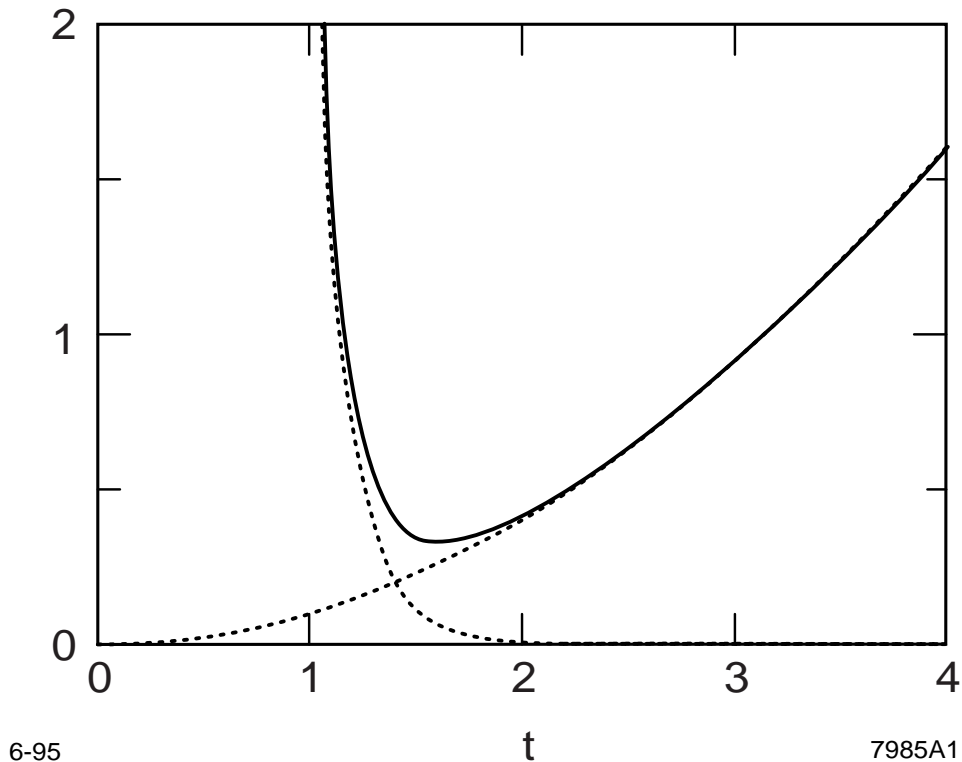}}
\end{center}
 \caption{The potential $V(t)$ for softly broken supersymmetric
 Yang-Mills theory with $\Nc = 3$, $\Nf = 2$.}
\label{Vless}\end{figure}

Thus, we may set $t_i = t$ for all $i$.  This gives
the expression
\beq
     V + \Delta V =  B_T\mQ \Nf |t|^2 + {1\over A_T}{\Nf
     \over |t|^{2\Nc/(\Nc-\Nf)}} .
\eeq{Vredt}
 It is easy to see that this expression is minimized for
\beq
     |t| = t_* = \biggl[{\Nc\over (\Nc - \Nf)} {1\over B_T A_T \mQ}
        \biggr]^{(\Nc - \Nf) \over 2 (2\Nc - \Nf) }
\eeq{tstardef}
The potential $V(t)$ is shown for the case $\Nf = 2$, $\Nc = 3$ in
Figure \ref{Vless}.

   The minimum of the potential can be brought by global symmetry
transformations into the form
\beq
        \VEV{T} =   t_*\cdot \One \ ,
 \eeq{unitform}
where $\One$ is the unit matrix.  This  expectation value spontaneously
breaks \leqn{globalsymm} to $SU(\Nf)_V \times U(1)_B$.

\subsection{The Spectrum of the $R$ model}

It is straightforward to work out the spectrum of the model by
expanding about the minimum of $V$.  Consider first the bosons of
the model. A general  $\Nf \times \Nf$
 complex matrix $T$ can be parametrized in terms of real-valued
 component fields as
\beq
      T =  t_* e^{(t_V + i t_A)/\sqrt{2\Nf}} V U \ ,
\eeq{genT}
where
\beq
       V = e^{t_{Vi}\lambda^i} \ ; \qquad U = e^{it_{Ai}\lambda^i} \ ,
\eeq{VandUdefs}
and the $\lambda^i$ are $SU(\Nf)$ matrices, normalized to
 $\tr[\lambda^i \lambda^j] = \half\delta^{ij}$.  In this
 parametrization, $|\det T|$ is a function of $t_V$ only, and the
 various
 real-valued components all  have kinetic energy terms of the form
\beq
        \L = \sum_I  \half A_T (\del_\mu t_I)^2  + \cdots .
\eeq{kincomps}

The fields $t_{Ai}$ and $t_{A}$ drop out of the potential completely.  This
is natural, because they are the Goldstone bosons of the spontaneously
broken $SU(\Nf)$ and $U(1)_R$ symmetries.  The fields $t_{Vi}$, which
form an adjoint representation of the unbroken $SU(\Nf)$ flavor group,
obtain the mass
\beq
     m^2_{Vi} = \bigl({2\Nc-\Nf\over \Nc-\Nf}\bigr) {2\over A_T^2}
                       \bigl({1\over t_*} \bigr)^{2\Nc/(\Nc- \Nf)}\ ,
\eeq{mimass}
and the singlet field $t_V$ obtains the mass
\beq
    m^2_{V} = \bigl({\Nc^2-\Nc\Nf+\Nf^2\over(\Nc-\Nf)^2}\bigr) {2\over A_T^2}
                       \bigl({1\over t_*} \bigr)^{2\Nc/(\Nc- \Nf)}\ .
\eeq{mmmass}

The fermion masses can be read directly from the superpotential
\leqn{ADSpot}.  Expanding this formula about the minimum according to
\beq
     T = t_* \One + \theta \cdot \Bigl( \psi_{T} {1\over \sqrt{2\Nf}} +
                \psi_{Ti} \lambda^i \Bigr) + \cdots ,
 \eeq{genTf}
we find mass terms for the flavor-singlet and -adjoint fermions:
\beqa
   m_\psi &=& {\Nc \over (\Nc - \Nf)}{1\over A_T}\bigl({1\over
   t_*}\bigr)^{(2\Nc - \Nf)/(\Nc-\Nf)}\ , \CR
   m_{\psi i} &=& {1\over A_T}\bigl({1\over
   t_*}\bigr)^{(2\Nc - \Nf)/(\Nc-\Nf)}\ .
\eeqa{fmasses}
No fermions remain massless.

\subsection{The $\notR$ Model}

Now we introduce the more general supersymmetry breaking term with $\mg$
nonzero.  Though it is possible to discuss this term from the beginning
with $\mQ$ and $\mg$ treated on the same footing, it is simpler---and
one obtains qualitatively the same results---if we treat $\mg$ as a
perturbation on the $R$ model just described.

The superpotential term involving $\mg$ requires $\VEV{S}$.
Quite generally, we can obtain the expectation value of $S$ from the
superpotential of a
supersymmetric effective Lagrangian by using the formula
\beq
    \VEV{S} =  {\del\over \del \log \Lambda^{(3\Nc - \Nf)}}  W \ .
\eeq{Sform}
This equation can be derived by starting from the effective Lagrangian
which includes $S$ explicitly \cite{VY,TVY}, or directly from
considerations of anomalies \cite{ILS}.

Restoring $\Lambda$ to
\leqn{ADSpot} and applying \leqn{Sform}, we find for the supersymmetry
breaking potential
\beq
    -\mg S =   - {\mg\over (\det T)^{1/(\Nc-\Nf)}}
   \  .
\eeq{Sforlow}
This potential depends on the phase of $\det T$, and thus it induces a
mass for the field $t_A$ in \leqn{genT}.  We find
\beq
   m_A^2 = {\Nf\over (\Nc-\Nf)^2 } {\mg \over A_T}\bigl({1\over t_*}
   \bigr)^{(2\Nc-\Nf)/(\Nc-\Nf)} .
\eeq{Amass}
The appearance of this mass term is expected:  The gluino mass term
explicitly breaks the $U(1)_R$ global symmetry and so should give mass
to the corresponding Goldstone boson.

  It is not difficult to work out the general formulae for the other
particle masses to first order in $\mQ$ and $\mg$.  However, there
are no surprises.  The vacuum remains unique up to global symmetry
transformations, and all of the particles except the $SU(\Nf)$ Goldstone
bosons remain massive.

  We can now discuss the extension of our results to more general forms
  for the K\"ahler potential.
  Because the spectrum we have found is the generic spectrum  for the
symmetry-breaking pattern we have observed, sufficiently small
perturbations of the K\"ahler potential do not affect the qualitative
physics.  It is possible to choose K\"ahler potentials which decrease
sufficiently strongly as the $t_i$ increase that the potential has more
than one minimum.  In this situation, it is formally possible
to have a minimum of $V$ in which the eigenvalues of $T$ take distinct
values.  In such a case, the vectorial flavor $SU(\Nf)$ symmetry is also
partially broken.  We do not consider this scenario likely, but
 we cannot rule
it out. Nevertheless, we will disregard this possibility in the rest of
our discussion.

 \subsection{Decoupling of Superpartners}

In the arguments just concluded, we have calculated the symmetry
breaking pattern and the spectrum of supersymmetric Yang Mills theory
perturbed to first order in soft supersymmetry breaking terms.
It is interesting that our results for the
global symmetry and the  massless
particles reproduce the standard expectations for chiral symmetry
breaking in   $\Nf$-flavor  QCD.
The final symmetry breaking pattern leaves a global symmetry
$SU(\Nf)_V\times U(1)_B$, and the only massless particles are the
Goldstone bosons corresponding to this symmetry breaking.
In QCD, this expectation is not particularly well supported for large
values of $\Nf$, but it is known to hold in the case which has been studied
experimentally, $\Nc = 3$, $\Nf = 2$, and in the limit $\Nc \ra
\infty$, $\Nf$ fixed \cite{CW}.

Thus, we feel confident in conjecturing that
the results we have obtained, at
first order in supersymmetry breaking, are smoothly connected to
the limit $\mQ, \mg \ra \infty$, in which the superpartners decouple
and the system reverts to an  ordinary Yang-Mills theory with fermions.
It is reasonable that this smooth extrapolation should apply quite
generally for $\Nf < \Nc$.  We will need to explore case by case
whether a similar extrapolation can hold for larger numbers of flavors.

There are two features of this extrapolation which deserve further
comment.  First,  in QCD, chiral symmetry breaking is characterized by
a nonzero vacuum expectation value of the quark-antiquark bilinear,
 $\psi_Q^i\psi_{\Qb j}$ in our present notation.  In the language of
the supersymmetric effective Lagrangian, this operator is a part of the
$F$ term of the superfield $T^i{}_j$.  The expectation value of this term
may easily be found to be proportional to
\beq
t_*^{-{N_c\over{N_c-N_f}}} \sim m_Q^{N_c\over{2N_c-N_f}} \ .
\eeq{Festimate}
Thus, the $F$ term of $T$ does obtain an expectation value in the
vacuum state that we have found. This expectation value naturally
becomes a nonzero expectation value for the quark bilinear in the
decoupling limit. As $m_Q$ increases, the quark bilinear
becomes larger while the squark bilinear becomes smaller, in exact
accord with our expectations.

When $\mQ$ is small, the vacuum we have identified
occurs at a very large value of $\VEV{T}$.  When $\VEV{T}$ is large,
the behavior of supersymmetric Yang-Mills theory can be described
classically, as the spontaneous breaking of the $SU(\Nc)$ gauge
symmetry by squark field vacuum expectation values.  In other words,
the gauge symmetry is realized in the Higgs phase.  However, since the
matter fields belong to the fundamental representation, there is no
invariant distinction between the Higgs and confinement phases of this
model, and so there is no impediment to the Higgs phase at small $\mQ$
being smoothly connected to a confinement phase at large $\mQ$.

\section{$\Nf =  \Nc$}

In the case $\Nf < \Nc$, we have found a very natural connection
between the physics of the theory with weak supersymmetry breaking and
the physics of the theory after the supersymmetric partners have been
decoupled.  For larger numbers of flavors, however, this connection
will become increasingly tenuous.

We next consider the case $\Nf = \Nc$.  Here the
low-energy effective Lagrangian of the supersymmetric limit contains
both meson and baryon superfields.  In this special situation, the
baryon fields $B$, $\Bb$ are flavor singlets, and both the meson
fields $T^i{}_j$ and the baryon fields have zero $R$ charge.
Seiberg has argued \cite{NatiMod}
that this model has a manifold of supersymmetric
ground states, in which the meson and baryon fields satisfy the
relation (in units where $\Lambda = 1$)
\beq
      \det T  -   B \Bb =   1 \ .
\eeq{Sconstraint}
Many forms for the superpotential are consistent with this relation.
The $S$-dependent superpotential, for example, has the form
\beq
     W  =   S \log \bigl(\det T - B\Bb \bigr) .
\eeq{Sdepconstraint}
Note that this superpotential leads to conditions for a
supersymmetric vacuum state which imply not only \leqn{Sconstraint}
but also the constraint $\VEV{S} = 0$, so that the $U(1)_R$ symmetry is
not spontaneously broken.

\subsection{Location of the Vacuum States}

The presence of a  manifold of degenerate vacuum states not related
by a global symmetry is necessarily accidental unless it is a
 a consequence of supersymmetry.  Thus, any such
degeneracy should be broken as soon as supersymmetry breaking terms are
added to the Lagrangian.  At first order, this is the main effect of the
soft supersymmetry breaking perturbation.  To analyze this effect, we
should restrict our attention to the values of
 $T$, $B$, and $\Bb$ obeying the constraint
\leqn{Sconstraint}, for which the vacuum energy vanishes in the
supersymmetric limit, and study the behavior of the supersymmetry
breaking potential over this space.

For simplicity, we begin with the $R$ models, for which $\mg = 0$.
Then the soft supersymmetry breaking terms \leqn{SSoft} lead to the
potential
\beq
  \Delta V =  B_T \mQ \tr \bigl[ T^\dagger T \bigr]  + B_B \mQ \bigl(
                          B^\dagger B + \Bb^\dagger \Bb \bigr) .
\eeq{delVwB}
Using $SU(\Nf)\times SU(\Nf)$, we can diagonalize $T$ to complex
eigenvalues $t_i$.  Parameterize the baryon fields as
\beq
            B = x b \ , \qquad   \Bb = - {1\over x}b \ ,
\eeq{Bparam}
with $x$ and $b$ complex. Then $b$ obeys the constraint
\beq
            \prod_i t_i + b^2 = 1 .
\eeq{tbconst}
The variable $x$ appears in the potential only through the baryon
mass term
\beq
   \Delta V =  \cdots +  B_B\mQ \bigl( \bigl|x\bigr|^2
   + \bigl|{1\over x}\bigr|^2 \bigr) |b|^2 \ ,
\eeq{xdep}
and this is minimized at $|x|= 1$ for any $b$.  Thus, we may set $|x| =
1$.

The problem becomes that of minimizing
\beq
    \Delta V = B_T \mQ \sum_i | t_i|^2 + 2B_B \mQ |b|^2
\eeq{newmineq}
subject to the constraint \leqn{tbconst}.  There are three types of
stationary points of this potential:

 (1)  If $b=0$, $\Delta V$ is stationary
when $|t_i|$ are all equal:
\beq
         |t_i| = 1 \ , \qquad   \prod_i t_i = 1\ ,\qquad b = 0 \ .
\eeq{tvac}

(2) If $T = 0$, $\Delta V$ is stationary:
\beq
        T = 0 \ , \qquad  b = \pm 1 \ .
\eeq{bvac}

(3) If neither $T$ nor $b$ vanish, there can be an additional stationary
point with $|t_i|^{(\Nf-2)} = (B_T/B_B)$ for all $i$.  This point is
always unstable with respect to the other vacuum states.

\begin{figure}
\begin{center}
\leavevmode
{\epsfxsize=5.00truein \epsfbox{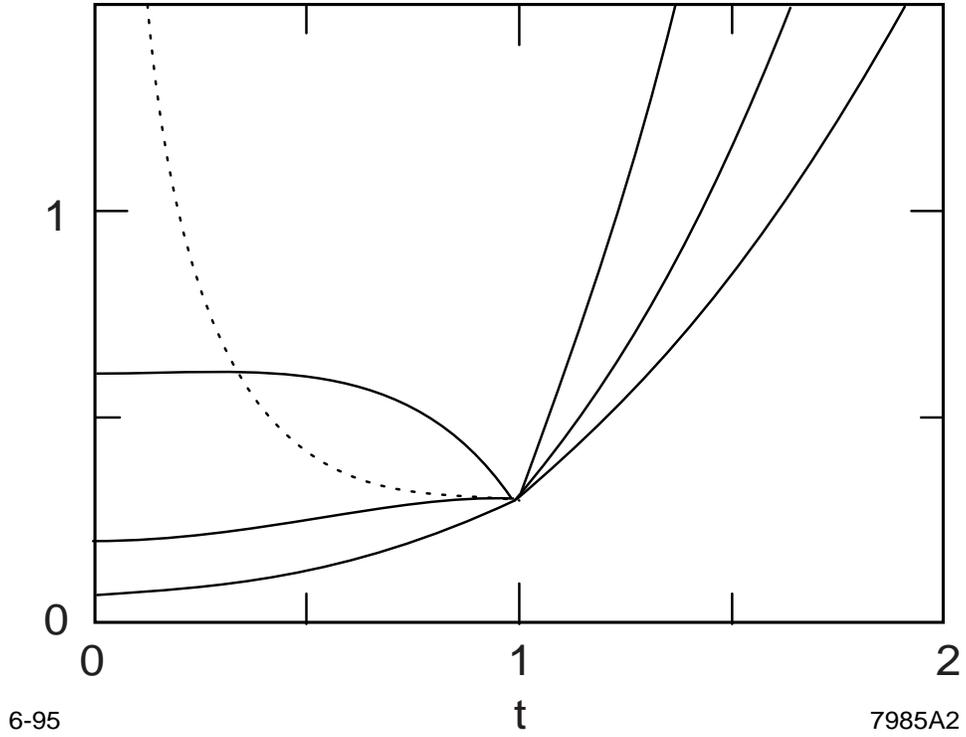}}
\end{center}
 \caption{The potential $\Delta V$ for softly broken supersymmetric
 Yang-Mills theory with $\Nc = 3$, $\Nf = 3$.  The potential is shown
 on the subspace $T = t \cdot \One$, as a function of $t$.  The three
 curves correspond to $(B_B/B_T) = {1\over 3}, 1, 3$, from bottom to top.
 The dotted line shows the location of the stationary point (3)
 referred to in the text.}
\label{Vequal}\end{figure}

The shape of the potential $\Delta V$, for three choices of
$(B_T/B_B)$,
is shown in Figure \ref{Vequal}.  Notice that the vacuum at $b=0$ is
the absolute minimum for sufficiently large values of $(B_B/B_T)$, but that
the vacuum at $T = 0$ is always a local minimum.


The method of effective Lagrangians cannot tell us which of the two
vacuum states at $b=0$ and $T=0$ is the preferred one.  This depends on
the ratio  $B_B/B_T$, which is a phenomenological input to the
effective Lagrangian analysis.  We will see below that the vacuum at
$T=0$ is locally stable if $B_B > B_T$ and is globally stable if
$B_B > (\Nf/2)B_T$.  In \leqn{BBTratio}, we attempted to estimate the
ratio of $B_B$ and $B_T$. Our naive estimate puts the theory just at
the boundary at which the two vacuum states have equal energy.
Probably, this question can only be decided by computer simulations.
We note, however, that if the
vacuum structure of this theory were being studied in a lattice
simulation, one could bias the simulation in favor of one vacuum or the
other by adding an explicit $B_B$ or $B_T$ term to the Lagrangian.
In the discussion to follow, we will treat each locally stable vacuum
state as if it could be separately  realized in a such a computer
experiment.

Up to now, we have ignored the possible effects of the
$U(1)_R$-violating supersymmetry breaking term proportional to $\mg$.
However, these effects cannot change the qualitative picture when
$\mg$ is small.  We showed earlier that
 the superpotential \leqn{Sdepconstraint} implies that,
in the manifold of
supersymmetric vacuum states about which we are perturbing, $\VEV{S} =
0$.  Thus, the superpotential term proportional to $\Mg$ does not
contribute to the vacuum energy.  More generally, since $\Mg$, $T$,
$B$, and $\Bb$ are all invariant under $U(1)_R$, while a superpotential
has $R$ charge 2, this term does not contribute to the superpotential
to any order in $\mg$.  There are possible K\"ahler potential terms
involving $\Mg$. (The simplest one will be discussed in a moment.)
However, near the vacuum with $b= 0$, these will be polynomials in
$B$ and $\Bb$ of order at least 2, and near the vacuum with $T=0$ they
will be polynomials in $T$ of order at least 2.  Thus, these terms will
not affect the presence of stationary points of the vacuum energy at
these positions in the field space.  These terms may alter the details
of the mass spectrum computed below, but they will not alter the
qualitative physical picture of vacuum stability which follows from
this calculation.

\subsection{The Spectrum at $b = 0$}

We will now work out the spectrum of particle masses at the two
candidate vacuum states that we have identified.  The boson masses can
be found by expanding the potential \leqn{delVwB} about the two
vacuum states, with fields subject to the constraint \leqn{Sconstraint}.
At this level, the fermionic partners of these fields remain massless.
Fermion masses will be induced when we include effects of first order in
$\mg$.

To expand about the vacuum at $b = 0$, parameterize $T$ as in
\leqn{genT}, with $t_* = 1$, and parameterize
\beq
        B = b + c \ ,  \qquad      \Bb =   - (b -c ) \ .
\eeq{BBbar}
The complex fields $b$ and $c$ have kinetic energy terms proportional
to the factor $(2 A_B)$, which must be divided out in computing masses.
The fields $t_V$ and $t_A$ in \leqn{genT} are removed by the constraint.
To leading order, \leqn{Sconstraint} implies
\beq
      t_V + i t_A = -  \sqrt{{2\over \Nf}} (b^2 - c^2) \ ,
\eeq{tVexpand}
to quadratic order in baryon fields.  Now we simply expand $\Delta V$
and read off the spectrum of masses.  We find, respectively for the
masses of $t_{Vi}$, the real part of $b$ and the imaginary part of $c$,
and the imaginary part of $b$ and the real part of $c$,
\beqa
     m^2_{Vi} & = &   {2\over A_T} B_T \mQ   \CR
     m^2_-    & = & {2\over A_B} (B_B - B_T) \mQ \CR
     m^2_+    & = & {2\over A_B} (B_B + B_T) \mQ \ .
\eeqa{equalNmasses}

Notice that, just at $B_B = B_T$, when the unstable stationary point (3)
 interposes
itself between the $b=0$ and $T= 0$ vacuum states, the $b =0$ vacuum
becomes locally stable with respect to baryon-number violating fields.  Since
the energies of the $b=0$ and $T=0$ vacua are $(\Nf B_T \mQ)$
and $(2 B_B \mQ)$, respectively, the vacuum at $T=0$ remains the global
minimum of the potential as long as
\beq
              B_B < {\Nf\over 2} B_T \ .
\eeq{CCondition}
as we claimed at the end of the previous section.

The expectation value of $T$ in this vacuum spontaneously breaks
$SU(\Nf)\times SU(\Nf)$ to $SU(\Nf)_V$.  The fields $t_{Ai}$, which are
the Goldstone bosons corresponding to this symmetry breaking, remain at
zero mass.  Since $U(1)_R$ is not spontaneously broken, we expect no
singlet Goldstone boson in the spectrum, and, indeed, none appears.

Since  \leqn{Sconstraint} is a superfield constraint, it also
removes one fermion from the theory, specifically, the fermionic
partner of $\tr[T]$.  The other
fermionic components of $T$, $B$, and $\Bb$ remain
at zero mass at this level of the analysis.
  This is natural, because the mass terms for these fields
violate $U(1)_R$ by 2 units.  Thus, these mass terms can only be
induced when the $R$-charge breaking term proportional to $\mg$ is
added.  We have noted above that this term cannot induce a
superpotential.  However, it can induce a $D$-term contribution of the
form indicated in \leqn{SSoft}. There are many possibilities for such a
term; a set of simple examples is given by
\beq
   \Delta \L = \int d^4 \theta \bar{\Mg} \bigl( C_T \det T + C_B
   B\Bb\bigr)  + \hc \ ,
\eeq{FirstCtry}
where $C_T$ and $C_B$ are some constants.  If one begins from the
effective Lagrangian including $S$, with the canonical superpotential
and K\"ahler terms,
\beq
   \L = \int d^4 \theta  S^* S + \int d^2\theta \bigl( S\log(\det T -
   B\Bb) + \Mg S \bigr) + \hc \ ,
\eeq{fullSLag}
and integrates out $S$, one finds
\beq
   \Delta \L = \int d^4 \theta \bar{\Mg} \log(\det T - B\Bb)  + \hc ,
\eeq{SecondCtry}
which gives qualitatively similar results.  In the following
discussion, we will work with \leqn{FirstCtry}.

To obtain baryon masses from \leqn{FirstCtry}, expand the superfields
about the vacuum state $T = \One$, $b= 0$ according to \leqn{genTf} and
\beq
     B =  \theta \cdot \psi_B \ , \qquad   \Bb = \theta \cdot
     \psi_{\Bb} \ .
\eeq{genBB}
We find, for the flavor adjoint and baryonic fermions,
the  masses
\beqa
     m_{\psi i } & = & \half {C_T\over A_T} \mg  \CR
     m_{\psi B} & = & {C_B \over A_B} \mg \ .
\eeqa{equalfmasses}
No zero-mass fermions remain.

\subsection{The Spectrum at $T = 0$}

Using the same techniques, we can work out the spectrum of masses in the
vacuum at $T=0$.  For the scalars, parameterize $B$ and $\Bb$ by
\beq
      B = (1 + b + c ) \ ,   \Bb = - (1 + b - c ) \ .
\eeq{BBbartwo}
The constraint \leqn{Sconstraint} allows us to eliminate $b$:
\beq
      b = - \half \det T  + \half c^2 \ .
\eeq{belim}
The  contribution from $T$ is higher order than quadratic and so does
not affect the mass spectrum.  Inserting \leqn{BBbartwo} and \leqn{belim}
back into $\Delta V$ and expanding to quadratic order, we find the
following masses for the components of $T$ and the real part of $c$:
\beq
      m_T^2 = {B_T\over A_T} \mQ     \ ,   \qquad
       m_{cR}^2 = {B_B\over A_B} \mQ \ .
 \eeq{vacatBmasses}
The imaginary part of $c$ remains at zero mass, which is expected,
because this field is the Goldstone boson of spontaneously broken
baryon number symmetry $U(1)_B$.

The constraint \leqn{Sconstraint} removes one linear combination of the
fermionic components of the baryon fields.  Otherwise,
no fermion masses appear until we add the $R$-symmetry breaking terms
involving $\mg$.  Then the term \leqn{FirstCtry} gives mass to the
remaining baryonic fermion,
\beq
      m_{\psi B}  =  {2C_B \over A_B} \mg \ ,
\eeq{equalfinT}
but it leaves the fermionic components of $T$ massless.

 \subsection{Toward the Decoupling Limit}

In the vacuum state at $b=0$, when we include a nonzero gluino mass
$\mg$, we find again the standard symmetry breaking pattern expected in
QCD.  The global group $SU(\Nf)\times SU(\Nf) \times U(1)_B$ is
broken spontaneously to $SU(\Nf)_V\times U(1)_B$,
leaving no massless particles except
for the required Goldstone bosons.  It is reasonable to expect that
here, as in the cases considered in section 3, there is a smooth
transition from the situation of weak supersymmetry breaking
 to the decoupling limit $\mQ, \mg \ra \infty$.  The
symmetry breaking term \leqn{FirstCtry} also induces a nonzero $F$ term
for the $SU(\Nf)_V$ singlet part of $T$.
This term should go naturally, in the decoupling limit, into the
chiral symmetry breaking expectation value of the quark-antiquark
bilinear.

However, all of the other vacuum states that we have identified are
unusual and unexpected.  All of them contain massless composite
fermions.  The vacua at $T=0$ have restored chiral symmetry and
spontaneously broken baryon number.   Could these vacuum states survive
to large values of the supersymmetry breaking parameters?

To answer this question, we must first understand why these vacua
contain massless fermions.  In general, in a strongly-coupled gauge
theory, chiral symmetries with nonzero anomalies generate sum rules
over the spectrum of zero mass particles.  These sum rules can be
saturated either by Goldstone bosons, if one of the symmetries is
spontaneously broken, or by massless composite fermions, if the
symmetries remain exact.  In the latter case, the anomalies computed from
the composite fermions must match the anomalies of the original
fermions; this is the 't Hooft anomaly matching
condition \cite{tHooft,BFSY}.

The three unusual vacua discussed in this section, the $b=0$ vacuum of
the $R$ model and the $T=0$ vacua of the $R$ and $\notR$ models, all
have unbroken anomalous chiral symmetries.  In all cases, the fermionic
content of the supersymmetric model is known to provide a solution to
the 't Hooft anomaly conditions associated with these symmetries
\cite{NatiMod,MeV}.  In fact, one might say that the fermions are
protected from obtaining masses  by the 't Hooft anomaly
conditions, because providing masses for a subset of the multiplet of
fermions would leave over a set of fermions which violates the 't Hooft
conditions and is therefore inconsistent, unless the chiral symmetry is
broken.

An interesting illustration of this argument is found
by comparing the spectra of massless fermions in the two vacuum states
at $T=0$ in the $R$ and $\notR$ models.  In the $R$ model, we have
massless fermions in the following representations of  the unbroken
symmetry group
$SU(\Nf)\times SU(\Nf) \times U(1)_R$:
\beq
     (\Nf, \bar{\Nf},-1) +  (1,1,-1)  \ ,
\eeq{forRreps}
corresponding to the fermions in $T$ and a linear combination of
the fermions in $B$ and $\Bb$.  Both
multiplets are necessary to satisfy the anomaly conditions involving
$U(1)_R$.  When $U(1)_R$ is broken explicitly by $\mg$, these
conditions no longer need to be satisfied, and so the baryonic fermions
can obtain mass. According to \leqn{equalfinT}, they do.

Because the massless composite fermions in these vacuum states
exist in order to satisfy the 't Hooft anomaly conditions, the
qualitative properties of these vacuum states are quite rigid.
We should recall that the $T=0$ vacuum and the
$b=0$ vacuum, for $B_T < B_B$, are locally stable minima of the energy
for sufficiently small $\mQ$; thus, there is a finite range of
$\mQ$ for which the pattern of symmetry breaking remains unchanged.
Given this pattern of symmetry breaking, the multiplet of composite
fermions cannot obtain mass.  Even if the composite fermions contain
as constituents bosons $Q$ or $\Qb$ which obtain mass from the
$\mQ$ term, the composites are bound rigidly to remain at zero mass.
This idea, that composites of massive constituents may be forced
to remain massless in order to satisfy the 't Hooft condition, was
formulated by Preskill and Weinberg many years ago \cite{Weinberg}.

Even if a vacuum with unbroken chiral symmetry is globally unstable to
tunnelling processes, the 't Hooft argument applies as long as it is
locally stable.  Thus, a vacuum with unbroken chiral symmetry can only
disappear, as $\mQ,\mg \ra \infty$, through a second-order phase
transition.

With this introduction, we can speculate on the evolution of these
vacuum states as $\mQ$ and $\mg$ are increased from zero.  Consider
first the $b=0$ vacuum of the $R$ model.  As $\mQ$ is taken to
infinity, the squarks decouple, and the model becomes a purely fermionic
Yang-Mills theory with $\Nf$ quark flavors plus one fermion flavor
in the adjoint representation of the gauge group.
For small values of the
supersymmetry breaking mass $\mQ$,  this vacuum contains massless
fermions corresponding to the fermionic components of the superfields
$T$, $B$, and $\Bb$.  We might think of these as being built out of
scalars, with one squark replaced by a quark to give the composite spin
$\half$.  But it is also possible to build objects with the same
quantum numbers purely  out of fermions, by replacing
\beq
            Q^i \ra  \lambda^\alpha \psi_{Q \alpha}^i\ , \qquad
               \Qb_j \ra  \lambda^\alpha \psi_{\Qb \alpha j} \ ,
\eeq{Qreplace}
where $\alpha$ is a two-component spinor index and the gauge
indices are implicit.
Notice that this combination has the same quantum numbers as the
squark, including zero $R$ charge.  Then, for
example, the fermion created by $T^i{}_j$ could be constructed as
\beq
       \psi_{T\alpha} {}^i{}_j \ra  \psi_{Q \alpha}^i \lambda^\beta
                   \psi_{\Qb\beta  j} \ .
\eeq{fmesonform}
With this replacement, the composite fermions are built only out of
constituents which remain massless as the squarks are decoupled.  Thus,
it is apriori reasonable that the $b = 0$ vacuum of the $R$ model
could go smoothly into a vacuum of the  purely fermionic
Yang-Mills theory described above.  This vacuum would have broken
$SU(\Nf)\times SU(\Nf)$ but unbroken chiral $U(1)_R$, zero values for
the vacuum expectation values of quark-antiquark bilinears,
massless composite fermions in the adjoint representation
of flavor $SU(\Nf)_V$, and massless baryons.
We will refer to this
scenario as `option 1'.  It will have analogues in the models to be
discussed later; however,
 these analogous phases will be less well motivated.
    It is easy to see that the replacement \leqn{Qreplace}
can formally
 be used to build composite fermions with only fermionic
constituents in any model with unbroken $R$ symmetry.

The other possibility for this model is that, after the squarks
decouple, the gluino fields pair-condense,
in a second-order phase transition at some value of $\mQ$,
and the nonzero value of
the condensate $\VEV{\lambda\cdot \lambda}$
spontaneously breaks $U(1)_R$.  In this
case, the physics would revert to the usual symmetry-breaking pattern
of QCD, and the composite fermions would become massive.
The gluino condensate would make itself felt only by providing an extra
$SU(\Nf)$-singlet  Goldstone boson.  We will refer to this scenario as
`option 2'.

One way to understand the physical distinction between options 1 and 2
is to consider a question raised some time ago but never answered in a
satisfactory way:  If a fermion in a large color representation is
added to QCD, does its pair condensation to chiral symmetry breaking
occur at the usual QCD scale, or at much shorter distances
\cite{Marciano,Kogut}.
If gluinos pair-condense at very short distances, before
normal quarks feel the full forces of the strong interactions, then
option 2 would be favored.  If gluinos feel strong interactions at
more or less the same scale as quarks, option 1 is a reasonable
possibility.  An extreme model in which gluinos
condense and decouple at a very high scale, as suggested in the papers
just cited, appears  unlikely as a result of our analysis, because
we know that option 1 is actually realized when squarks are added back
to the model.

Consider next the vacuum state at $T=0$, first in the $\notR$ case.
In this model, the massless composite fermions belong to the
$(\Nf,\bar{\Nf})$ representation of an unbroken flavor group $SU(\Nf)\times
SU(\Nf)$. There are no constraints from the $U(1)_R$ symmetry, which is
explicitly broken, or from baryon number, which is spontaneously
broken.  With this freedom,
can we build these fermionic composites out of fields that survive in
the decoupling limit $\mQ, \mg \ra \infty$?  For $\Nc$ even, it is
impossible, because the only constituents available are the quarks
$\psi_Q^i$, $\psi_{\Qb j}$, and gauge-invariant states must contain
an even number of these.  For $\Nc$ odd,
however, it is possible to build
composites with the correct quantum numbers, as follows:
\beq
       \psi_{T\alpha} {}^i{}_j \ra  \epsilon^{ab \cdots d}\,
      \psi_{Q\alpha  a}^i\,  \epsilon_{jk\cdots \ell}
         \bar{\psi}^{\dot{\beta} k}_{\Qb b } \cdots
          \bar{\psi}^{\ell}_{\Qb \dot{\gamma} d}  \ ,
\eeq{Tmesonform}
where $\bar\psi_{\Qb}$ is the right-handed fermion field in
$(\Qb)^*$. The $(\Nc -1)$ right-handed fermion fields must be
contracted into a Lorentz scalar combination.  For the case $\Nf = \Nc
= 3$, eight of the nine fermions in \leqn{Tmesonform} have the
quantum numbers of the baryon octet in QCD.

However, in this case, there are two compelling arguments that the
spectrum which we find cannot survive to the decoupling limit.  In the
limit $\mQ \ra \infty$, even without  introducing $\mg$, we have a
vectorlike gauge theory of fermions.  For such theories, the QCD
inequalities of Weingarten \cite{Weingarten} and Vafa and Witten
\cite{VafaWitten} apply.  In the Appendix, we use Weingarten's method
to prove that, in the decoupling limit, flavor nonsinglet composite
fermions must be heavier than the pions, which are massive in the $T=0$
vacuum.  Alternatively, we can apply the theorem of Vafa and Witten in
the decoupling limit to show that
vectorlike global symmetries, in particular, baryon number, cannot
be spontaneously broken.

By either argument, the $T=0$ vacuum state
must disappear in a second order phase transition at a finite value of
$\mQ$.
  Most likely, this vacuum becomes
locally unstable  with respect to a decrease in
the expectation value of $b$, driving the theory back to the
more familiar vacuum  at $b=0$.

Finally, we may consider the $T=0$ vacuum in the
$R$ models.  The arguments that we have just
presented for the $T=0$ vacuum in the
$\notR$ models apply equally well to the $R$ case.  Again, we must
 have a second-order transition,  probably with an
instability to the $b =0$ vacuum.  There are then two possible
endpoints, depending on which option is chosen for the $b=0$ vacuum.
If the option 1 for the $b=0$
vacuum is correct, it is not necessary that $U(1)_R$ be spontaneously broken
in this transition.

\section{$N_f = ( N_c +1)$}

So far we have considered separately models with  $\Nf < \Nc$ and
models with $\Nf = \Nc$.
The cases where the number of flavors exceeds the number of colors fall into
two classes, those of $\Nf = \Nc+1$ and those of $\Nf > \Nc+1$.  These
two classes of theories have qualitatively similar physics, in both
cases much simpler than that of $\Nf = \Nc$.

In the case of $\Nf = \Nc+1$, like in  the case of $\Nf = \Nc$, the low-energy
effective Lagrangian  in the supersymmetric limit is expressed  in terms  of
the
baryon, anti-baryon and meson superfields. However, now these superfields
are not $R$ neutral, and the baryons are not flavor singlets.
Rather, they
transform in the  representations of the global symmetry
\leqn{globalsymm}
\beq
B\ : \  (\bar\Nf,1)_{1,1-{1\over \Nf}}\qquad \bar B:\ (1,\Nf)_{-1,1-{1\over
\Nf}}
 \qquad  T :\  (\Nf,\bar\Nf)_{0,{2\over \Nf}}
\eeq{NfNc1sym}
where the second subscript is the
$R$ charge of the scalar component of the superfield.
In the supersymmetric theory
the low energy effective theory is described (at least near the origin
of moduli space) by
the K\"ahler potential given by \leqn{cKahler},
and by the following  superpotential \cite{NatiMod}:
\beq
W = B_i T^i_{j}\bar B^{j} - \det T.
\eeq{W5}
The supersymmetric  vacuum is, thus,   described by a moduli
space characterized by
\beq
B_i T^i_{j} =\ 0\qquad  T^j_{i} {\bar B}^{ i}
 =\ 0\qquad {1\over N_c!}\epsilon_{i_1,...,i_{N_f}}
\epsilon^{ j_1,..., j_{N_f}} T^{i_1}_{ j_1}...
T^{i_{N_c}}_{ j_{N_c}} - B_{i_{N_f}} \bar B^{ j_{N_f}}=0.
\eeq{modspa}
As was argued in \cite{NatiMod}, these equations
correctly
describe the moduli space of vacuum states in the full
quantum theory.
At     the origin  of the moduli space,
 $<T> = <B>=<\Bb>=0$,  where the full global
symmetry \leqn{globalsymm} remains unbroken,
there  is a further consistency check for the low energy behavior.
The  fermionic components of the low-energy superfields \leqn{NfNc1sym}
match the
global anomalies of the underlying theory.

\subsection{The Vacuum}

When we break supersymmetry by squark and gluino masses, we add to the
effective Lagrangian the mass terms for $T$, $B$ and $\Bb$ indicated in
\leqn{SSoft}.  Since we are adding terms to the potential which are
positive  and vanish at the origin of moduli space, it is
obvious that the origin becomes the only vacuum state of the theory.
All of the scalar particles in the effective theory obtain mass
terms proportional to $B_Tm_Q^2$ or $B_B m_Q^2$.

Though all of the scalars obtain mass, all of the fermions remain
massless.  The superpotential \leqn{W5} is a least
cubic in fields, so any mass term derived from this superpotential
vanishes at the origin.  Similarly, in the $\notR$ case,
the $M_g$ term in \leqn{SSoft}
requires a function of $T$, $B$, and $\Bb$ which is neutral with
respect to the global group; the only such functions quadratic in
fields are $\tr[T^\dagger T]$, $B^\dagger B$, and $\Bb^\dagger \Bb$,
and these do not give fermion masses when integrated with $M_g$.
In fact, it is required that no fermions should obtain mass, since the
full multiplet of fermions in $T$, $B$, and $\Bb$ is needed to
satisfy the `t Hooft anomaly conditions for the remaining
global symmetry group
$SU(\Nf)\times SU(\Nf) \times U(1)_B$.

\subsection{Toward the Decoupling Limit}

The analysis of the previous section indicates that  in a finite region
of  small $m_Q$ and $m_g$, the ground state of the theory is a smooth
continuation of the origin of the moduli space of supersymmetric vacuum
states.  In this region of soft supersymmetry breaking parameters,
chiral symmetry is unbroken, and the full complement of fermions is kept
massless by the requirement that the 't Hooft anomaly conditions be
satisfied.  In the $\notR$ case, since both gluinos and squarks are
massive, at least some of the massless composite fermions must have
massive constituents.  As in our earlier examples, these particles are
protected from receiving mass by the 't Hooft conditions.

  In neither the $\notR$ nor the $R$ case, however, can this spectrum
of particles be correct in the decoupling limit.  In that limit, the
Weingarten inequality proved in the Appendix prohibits a
composite fermion which is nonsinglet in flavor from remaining massless
while the pion is massive.  In both cases, then, the phase we have
found at small $m_Q$ must disappear at a second-order phase transition
when $m_Q$ reaches a critical value.  In the $\notR$ case, the theory
has no option but to revert to the conventional pattern of symmetry
breaking in which the chiral symmetry group is broken to $SU(\Nf)_V\times
U(1)_B$ and all fermions become massive.

  For the $R$ case, however, there are still two options, corresponding
to options 1 and 2 described in Section 4.4.  Option 2 is the
scenario just described for the $\notR$ case, with symmetry breaking
to $SU(\Nf)_V\times U(1)_B$ and one extra Goldstone boson.  Option 1 is
the breakdown of the chiral symmetry group only to $SU(\Nf)_V\times
U(1)_B \times U(1)_R$.  In order to satisfy the 't Hooft anomaly
conditions associated with the $U(1)_R$, all of the fermionic
components of $T$, $B$, and $\Bb$ must remain massless.  As in the
case considered in Section 4.4, we can build all of the required
massless fermions out of quarks and gluinos by using the replacement
\leqn{Qreplace}.  In this case, as opposed to that of Section
4.4, the partial symmetry breaking required in option 1 is not
particularly well motivated.  However, we have not been able to rule
it out as a possibility.  We should also note that, even if this case
is realized in the more conventional option 2, the case $\Nf = \Nc$
could be realized in option 1.  There is no theorem that, when one
quark becomes very heavy, fermions not containing that quark cannot
become massless.

\section{$\Nf \geq (\Nc+2)$}

No solution of the 't Hooft anomaly matching conditions for
SQCD
involving gauge invariant
bound states is known for $N_f > (\Nc +1)$.  However, Seiberg has
suggested a compelling solution to these constraints in terms of new
gauge degrees of freedom which are dual to the original quarks and
gluons \cite{NatiDual}.  In this picture,
the theory is equivalent in the infrared to an SQCD theory
 with gauge group  $SU(N_f-N_c)$,
 $N_f$ dual quark flavors, and additional singlet fields
$T^i{}_j$ identified with the mesons of the original theory. The original
SQCD theory is infrared-free for $N_f \geq 3N_c$, so that in that case the low
energy description of the theory is in terms of the original quarks and
gluons. For $N_f \leq {3\over 2} N_c$, the dual ``magnetic'' theory is
infrared-free, and then
 the low energy description of the theory should be in terms
of the dual quarks, gluons and the singlet meson fields. In the
intermediate range ${3\over 2}N_c < N_f < 3N_c$ both theories are
asymptotically free. Seiberg suggested that, in this region, the
theory has a non-trivial infrared fixed point, and the theory has
dual descriptions in the infrared as interacting gauge theories with
superconformal global symmetry.
 While
the origin of this dynamically generated gauge symmetry is still unclear,
there is ample evidence that Seiberg's description of the SQCD theory is
correct, and we will assume it throughout this section.

If we break supersymmetry by giving masses to some of the fields of
SQCD, the leading term of the beta function will change for distances
greater than the scale of the masses.  The long-distance
gauge theories will be asymptotically free in a larger range of $\Nf$,
for
$N_f <
{9\over 2}N_c$ after adding squark masses, and for $N_f < {11\over 2}N_c$
after adding squark and gluino masses.  Beyond the point where the
theory is asymptotically free,
 we expect the effects of adding soft
SUSY breaking mass terms to be trivial. The massless particles are expected
to be infrared free, and there is no reason for the chiral symmetry to break.
We will concentrate our analysis, then, on the cases of $\Nf$
relatively close to the boundary $(\Nc+2)$, where the original gauge
theory becomes strongly coupled and the dual description is appropriate
in the infrared.  In the
next subsection we will analyze the effect of adding soft SUSY
breaking mass terms on the dual description of the theory. In the
second
subsection we will discuss in what range of $\Nf$ we expect this dual
description to be relevant, and speculate on the infrared
 behavior of the theory
for different values of $\Nf$.

\subsection{The Spectrum and Vacuum of the Dual Theory}

Seiberg's dual description of SQCD has an $SU(\tNc)$ local gauge
symmetry, where $\tNc = (\Nf - \Nc)$ as in \leqn{Ntilde}.
The elementary fields in the dual theory are
an $SU(\tNc)$ super gauge multiplet, $N_f$ flavors of dual quarks
$q_i^a$ and anti--quarks $\bq^j_a$ in the fundamental and anti--fundamental
representations of $SU(\tNc)$, respectively, and meson fields $T^i{}_j$.
The quark fields are in the $(\bnf,1)$ representation of the $SU(N_f)\times
SU(N_f)$ flavor group, the anti--quark fields are in the $(1,N_f)$
representation and the meson fields are in the $(N_f,\bnf)$ representation.
It is useful to think that the dual quarks are obtained by dissociating
a baryon \leqn{baryonop} into $\tNc$ components, and that the new gauge
fields parameterize a constraint which gives these baryons as its
solutions.  Seiberg also requires a superpotential
\beq
	W = T^i{}_j q_i^a \bq^j_a
\eeq{dualsp}
so that the scalar potential, including the $F$ and $D$ terms, is
\beqa
	V(T,q,\bq) &=& {1\over A_T} (\dq)_i^a (\dbq)_a^j q_i^b \bq^j_b +
		{1\over A_q} \bigl( (\dT)^i{}_j (\dbq)^j_a T^i{}_k \bq^k_a +
			(\dT)^i{}_j (\dq)_i^a T^l{}_j q_l^a \bigr) \CR &+&
		{g^2\over 2} ((\dq)_i \tau^A q_i - (\dbq)^j \tau^A
			\bq^j)^2 \ ,
\eeqa{dualscp}
where $g$ is the $SU(\tNc)$ gauge coupling, and $\tau^A$ are the
$SU(N_f-N_c)$ generators. $A_T$ and $A_q$ are the coefficients of the
corresponding (canonical) kinetic terms.
This scalar potential has  a
moduli space of vacua, which includes the point $<T> =<q> =< \bq> = 0$ at which
the chiral symmetry is unbroken \cite{NatiDual}.

Now add squark masses to the theory. Their effect should be seen in the
effective Lagrangian, and we can represent it by applying the logic of
Section 2.3 to the dual theory.  That is, we should  add to the
effective Lagrangian of the dual theory the term
\beq
	\Delta V = B_T m_Q^2 \tr(\dT T) + B_q m_Q^2 (|q|^2 + |\bq|^2),
\eeq{deltav}
at least near the origin of moduli space. After we add this
perturbation, the only minimum of the potential is at $<T> = <q> =
<\bq> = 0$.
Thus, adding a squark mass leaves the theory in the phase in which the
chiral symmetry is unbroken. All scalars get masses (originating only from
$\Delta V$, since the original scalar potential is quartic in the fields),
while all fermions remain massless.
As in the original supersymmetric theory, this complement of massless
fermions has just the right quantum numbers to satisfy the `t Hooft
anomaly conditions for completely unbroken chiral symmetry.  Thus, our
picture of the effect of soft supersymmetry breaking in this case is just
the same as in the case $\Nf = (\Nc+1)$ considered in the previous
section, except that the baryons of that case are replaced here by
their constituent dual quarks.

The glueball operator $\tr(W_{\alpha}^2)$ is identified (up to a sign)
between the original and the dual theory \cite{NatiIntDual}.
Thus, to leading order in $m_g$,
a gluino mass in the original theory is just equal to a gluino mass in the
dual theory. Adding this term breaks the $U(1)_R$ symmetry, but the
$SU(N_f)\times SU(N_f)$ global symmetry still remains and protects the
dual quarks from getting a mass. Thus, we find the same spectrum in the $R$
and $\notR$ cases, except that in the latter case the dual gluino,
which can be an asymptotic particle, becomes massive.

\subsection{Toward the Decoupling Limit}

Let us discuss now the infrared description of the theory.
We consider first the case of small $m_Q$ (and small
 $m_g$, in the $\notR$ case).  We have already remarked that, for
 $\Nf > {11\over 2}\Nc$ ($\Nf > {9\over
2} \Nc$ in the $\notR$ case), the theory becomes free in the infrared
 and is well described in terms of the original variables---gluons and
 quarks (and gluinos in the $\notR$ case).  This statement applies
equally well to the dual version of the theory.  Thus, for  $\Nf >
{11\over 2} \tNc$, or $\Nf < {11\over 9}\Nc$, the dual theory is free
in the infrared.  For the $\notR$ theory, the corresponding criterion
is $\Nf < {9\over 7} \Nc$.  In this range of $\Nf$, the spectrum of the
theory contains massless dual quarks interacting weakly through a
dual gauge field which becomes asymptotically weak at large distances.
Unfortunately, this range of $\Nf$ is rather narrow; the first example
requires an $SU(8)$ gauge group and 10 flavors, even in the $\notR$
case.

However, it is likely that Seiberg's duality would still hold in the
 intermediate range of $\Nf$:  ${11\over 9}N_c < N_f < {11\over 2}N_c$
 in the $R$ case and
${9\over 7}N_c < N_f < {9\over 2}N_c$ in the $\notR$ case. As in the
supersymmetric case, we can prove the existence of an infrared fixed
point for values of $\Nf$ very close to the boundary of this region
by using the fact that the second coefficient of the QCD beta function
is positive when the first coefficient vanishes \cite{BandZ}.
Thus, some part of this intermediate range is controlled by a
nonsupersymmetric infrared fixed point.  At least when the fixed point
coupling is sufficiently small, the chiral symmetries of the theory
remain unbroken and the spectrum still  contains massless quarks or dual
quarks.  If at some value of $\Nf$, the massless fermions are no longer
asymptotic states, then also the solution to the `t Hooft anomaly
conditions is lost and the theory reverts to a scenario with broken
chiral symmetry.

The discussion of the decoupling limit for these theories is very
similar to that for the $\Nf = (\Nc+1)$ theory.  If the full chiral
symmetry group remains unbroken for small values of $m_Q$, the
 fermions in
the supermultiplet $T$ still
cannot remain massless in the decoupling limit where we have
the QCD inequality, precisely as discussed in Section 5.2.
Thus, those values of $\Nf$ which have massless fermions for small
values of $m_Q$ must have a second-order phase transition as $m_Q$ is
increased.  It is not clear how the theory behaves on the other side of
this phase transition. In the $\notR$ case obviously only a $SU(\Nf)_V\times
U(1)_B$ symmetry remains, with no massless fermions.
However, in the $R$ case, we can
use the dual fermions in $T$, $q$, and $\bq$ to solve the `t Hooft
anomaly equations associated with $U(1)_R$.  Thus, in this case, we
have available both option 1, in which the chiral group is broken to
$SU(\Nf)_V\times U(1)_B \times U(1)_R$ and all fermions remain massless,
and option 2, in which the chiral group is broken to $SU(\Nf)_V\times
U(1)_B$ and all fermions obtain mass.

\section{$\Nc = 2$}

For $N_c=2$ there is no distinction between massless
quarks and anti-quarks, so
that the global symmetry  in the supersymmetric limit
is $SU(2N_f)\times U(1)_R$
instead of $SU(N_f)\times SU(N_f)\times U(1)_B\times U(1)_R$. This changes
some
of the details in the discussions above, but does not change the qualitative
picture. The meson is now given by $T^{ij} = Q^i Q^j$, in the
anti--symmetric representation of $SU(2N_f)$, and the superpotential
generally involves ${\rm Pf}(T)$
instead of $\det(T)$. There are no baryon operators in this case;
rather, the baryons of the previous examples are absorbed into the
extended meson multiplet.
In the usual QCD theory with 2 colors, the global symmetry breaks from
$SU(2N_f)$ to $Sp(2N_f)$ (we denote by $Sp(2N_f)$ the $Sp$ group whose
fundamental representation is of size $2N_f$).
We shall now discuss the picture after soft SUSY breaking
for each relevant value of $N_f$.

For $\Nf = 1$, the
behavior is similar to the other cases $\Nf < \Nc$.  The
effective Lagrangian has a superpotential of the form
$W = 1/{\rm Pf}(T)$.  There is just one
vacuum, in which $T^{12}$ obtains an expectation value,
 breaking the flavor symmetry from
$SU(2)\times U(1)_R$ to $Sp(2)$ (which is isomorphic to $SU(2)$). The meson
$T^{12}$ is the goldstone boson for the breaking of the $U(1)_R$ symmetry
in the $R$ case; this particle
obtains a mass when we add a gluino mass.
A smooth transition is expected to the decoupling limit, as for $N_c > 2$.

For $\Nf = 2$, the moduli space of supersymmetric vacuum states is
constrained by the equation
${\rm Pf}(T) = 1$
\cite{NatiMod}. As in section 4, the potential from the soft
supersymmetry breaking terms can be considered on the space satisfying
this constraint.  Then, up to global symmetry
transformations, there is just one stable vacuum, for which
\beq
	T = \left( \begin{array}{cccc}
		   \sigma^2 & 0 \\
		   0 & \sigma^2
                   \end{array} \right) .
\eeq{TVev}
This breaks the $SU(4)$ flavor symmetry to $Sp(4)$.
In the $R$ case, the $U(1)_R$
symmetry is left intact.
 The fermionic fluctuations around the vacuum \leqn{TVev},
which transform as $6_1$ under $SU(4)\times U(1)_R$, decompose under
$Sp(4)\times U(1)_R$ as
$(5+1)_1$.
For small values of the SUSY breaking parameters,
the fermions in $5_1$ remain massless and
satisfy the 't Hooft anomaly matching conditions for the unbroken symmetry
group $Sp(4)\times U(1)_R$ \cite{NatiMod}.  In the $R$ case, there are
two options for the decoupling limit, as in the $b=0$ vacuum of Section
4.  In option 1, this spectrum continues smoothly to the decoupling
limit.  In option 2, the $U(1)_R$ symmetry is spontaneously broken and
the fermions in the $5_1$ obtain mass.  In the $\notR$ case, as in the
discussion of Section 4, the vacuum state obtained for small
supersymmetry breaking has no massless fermions and can smoothly become
the standard QCD vacuum as $m_Q\to \infty$.

For $\Nf = 3$, the
effective description of the SQCD theory has a
superpotential of the form $W = -{\rm Pf}(T)$ \cite{NatiMod}. In the
supersymmetric
case there is a moduli space of vacua, but adding the squark masses leaves
only the vacuum at $T=0$, as for $N_c > 2$. At this vacuum the chiral
symmetry is unbroken. All of the fermions, which are in the
$15_{-1/3}$ representation of the global symmetry,
 remain massless, and this multiplet satisfies the 't Hooft
anomaly matching constraints \cite{NatiMod}. As in section 5, in the
decoupling limit we expect a second order phase transition,
breaking the global symmetry from $SU(6)\times U(1)_R$ to $Sp(6)\times
U(1)_R$ or to $Sp(6)$.

For $\Nf \geq 4$,
the SQCD theory has a description in terms of dual gauge variables.
For $\Nf < 6$, the theory is conjectured to be described by an infrared
fixed point.
 As in section 6, we expect the
theory near the supersymmetric
 point  to be either at
some non-trivial infrared fixed point with the chiral symmetry unbroken, or to
be in a QCD-like phase in which the chiral symmetry breaks to $Sp(2N_f)$.
In  these cases,
the dual gauge group is  always asymptotically free, and so
we do not expect a phase in which the dual gauge symmetry is weakly
coupled.
Thus, the dynamically generated gauge symmetry
suggested by Seiberg should be difficult to identify
in simulations with the gauge group $SU(2)$.

\section{Problems of Approximate Supersymmetry on the Lattice}

Can the phenomena we have discussed in this paper be seen in lattice
gauge theory simulations?  Throughout this paper, we have considered
only soft supersymmetry breaking perturbations.  However, since, in
general, gauge
theories on the lattice cannot be made supersymmetric at the
fundamental level, we expect that lattice simulations of these theories
will also contain small dimension 4 perturbations which violate
supersymmetry.  Our analysis has been based on the assumption that, if
the phenomena discussed by Seiberg survive perturbations which are
relevant in the infrared, they should also survive small marginal
perturbations.

However, there is a serious difficulty with
this logic.   Our argument does not apply unless we can reach the
continuum limit.  But typically in lattice gauge theory simulations
with scalar fields, there is no continuum limit; instead, one finds a
first order phase transition as a function of the scalar field mass
parameter \cite{FraShe}.  This fact is
understood using the mechanism discovered by Coleman and Weinberg
\cite{CandWn}:  Renormalization effects in a gauge theory can
induce an unstable potential for a scalar field coupled to the gauge
bosons, leading to a `fluctuation-induced first-order phase
transition'.  We must ask whether there is a possibility of such
first-order phase transitions in approximately supersymmetric models,
and, if so, how they can be avoided.

To analyze this question, consider the renormalization group equations
for an approximately supersymmetric gauge theory.  Viewed as a
conventional renormalizable gauge theory, SQCD has three coupling
constants, the gauge coupling $g$, the quark-squark-gluino coupling
$g_\lambda$, and the four-scalar coupling $g_D$.  The scalar potential
has the  specific form
\beq
    V = { g_D^2\over 2} \biggl[ Q^\dagger \tau^A Q
    - \Qb \tau^A \Qb^\dagger
                \biggr]^2 \ ,
\eeq{Dpot}
where $\tau^A$ is an $SU(\Nc)$ matrix.
If we relax the constraint of supersymmetry, there are four possible
invariants under the symmetries of the problem, including the
continuous global symmetries and parity $Q \leftrightarrow
\Qb^\dagger$. The most general linear combination of these invariants
can be generated by the renormalization group flow.

We will view the lattice theory as providing a finite cutoff for the
quantum field theory, which does not respect supersymmetry. In this
cutoff field theory we will choose the bare couplings to obey the
supersymmetry relations, at least
approximately. In particular, we will choose the bare scalar
potential to be given by \leqn{Dpot}. The radiative corrections will cause
a finite renormalization of the couplings, which will violate supersymmetry
and generate other scalar potential terms. We expect the generated terms
to be smaller than the original terms. Our analysis of the
renormalization group flow of the theory will, therefore, be performed near
the
supersymmetric point. In particular we will restrict our analysis to the
surface given by
the three couplings $g$, $g_\lambda$, and $g_D$. We assume that our initial
conditions lie near this surface, and we are interested in the flow of the
couplings towards the infrared. Note that since we
we are not interested in scaling towards the continuum
limit, we do not analyze here the flow of the couplings towards the
ultraviolet. It is not possible to ensure that all couplings tend
smoothly to zero in the ultraviolet without fine adjustment of their
initial values.

In the surface given by $g$, $g_\lambda$ and $g_D$, the
beta functions of the three couplings are given (to leading order in
perturbation theory) by
\beqa
   \beta_g &=& -{1\over (4\pi)^2}\bigl[ 3\Nc - \Nf \bigr]g^3 \CR
   \beta_{g_\lambda} &=& -{1\over (4\pi)^2}\biggl[ g_\lambda g^2
      \bigl( 3\Nc + 3 C_2(\Nc)\bigr)
            - g_\lambda^3 \bigl( 3C_2(\Nc) +
            \Nf\bigr)\biggr] \CR
   \beta_{g_D^2} &=& - {1\over (4\pi)^2} \biggl[ 4g_\lambda^4 \Nc
                         + 2 g_D^4 \bigl(\Nc - \Nf - 2C_2(\Nc)\bigr)\CR
          & &\qquad \qquad \qquad
            + 12 g_D^2 g^2 C_2(\Nc) - 8 g_D^2 g_\lambda^2
                   C_2(\Nc) \biggr] \ ,
\eeqa{thebetas}
where $C_2(\Nc) = (\Nc^2-1)/2\Nc$.  These three functions all reduce to
the standard SQCD beta function on the supersymmetric subspace; for
 $g^2 = g_\lambda^2 = g_D^2$, $\beta_g = \beta_{g_\lambda} =
 \beta_{g^2_D}/2g$.  Note that, for $\Nc \sim \Nf $ and the three
 couplings in reasonable ratio, all three couplings are infrared
 unstable. In particular, $g_D^2$ is renormalized toward larger
 positive values.

 The potential instability to a first order phase transition
 arises because a new structure in the potential is induced by the
 renormalization group flow. To lowest order, the form of the potential
 induced is
 \beq
    V_E = { g_E^2\over 2} \biggl[ Q^\dagger
    \{\tau^A,\tau^B\} Q
    + \Qb \{\tau^A,\tau^B\} \Qb^\dagger
                \biggr]^2 \ ,
\eeq{Epot}
On the surface $g_E^2 = 0$, the beta function for $g_E^2$ is
 \beq
   \beta_{g_E^2} = - {1\over (4\pi)^2} \biggl[ 4g_\lambda^4
                              - 3 g^4 - g_D^4 \biggr] .
\eeq{Ebetas}
This equation implies that, if
one  leaves out the gluinos, $g_E^2$ becomes negative in the
 infrared, leading to  a fluctuation-induced first-order phase
 transition.  According to \leqn{Ebetas}, this effect is removed if the
 lattice simulation includes gluinos, and if the gluino coupling
 $g_\lambda$ is large enough. If we choose initial conditions in which
$g_\lambda$ is slightly larger than $g$, equation \leqn{thebetas}
guarantees
that this condition will be preserved along the renormalization group
flow. Equation \leqn{Ebetas} then shows that
no instability
is generated in the perturbative region.
Hopefully, this perturbative result remains valid
as we flow towards the infrared.

With this provision to avoid possible first-order phase transitions,
we expect that lattice simulations with an approximately
supersymmetric action can reach the continuum limit and test our
predictions for softly broken supersymmetric QCD.

\section{Summary and Conclusions}

In this paper we investigated softly broken $N=1$
supersymmetric QCD . We considered two types of soft
breaking terms, associated with squark masses $m_Q$
with or without additional
gluino masses
$m_g$. We denoted these cases by  $\notR$ and $R$, respectively.
 In the  limit  of $m_Q, m_g\rightarrow \infty $ the $\notR$ case
 should go over
to ordinary QCD, while in the $R$ case, in the limit $m_Q \rightarrow
\infty$, we recover QCD with an additional
massless adjoint fermion.
The two main questions that we addressed
are:
\begin{description}
\item[-]
To what extent do the results which were recently obtained for $N=1$ $SU(N_c)$
SQCD \cite{NatiMod,NatiDual}, and for other $N=1$ supersymmetric
gauge  theories as well,
carry over to the non--supersymmetric case?
  Is supersymmetry an essential prerequisite for those exotic
  phenomena?

\item[-]
How does the theory behave in the
decoupling limit, in which we take the soft breaking terms
($m_Q$ in the R case and $m_Q$, $m_g$
in the $\notR$ case) to be very
large compared to the dynamically generated scale $\Lambda$?
\end{description}

Our main results are the following :

 (i) {\it  All  the ``exotic'' phenomena that characterize the supersymmetric
theory continue to exist for  small values of the soft breaking mass
parameters. }

 It seems that the appearance of the exotic behavior
 is not related to
supersymmetry,
though it probably is related to the presence of fundamental
scalar fields.
Theories which include scalar fields generally do not possess a positive
definite measure for the gauge fields; this is the case in particular for
supersymmetric gauge theories as well as for the softly broken
supersymmetric theories. In these cases we cannot apply the QCD
inequalities method, as used in the appendix, to obtain information about
the theory. We recall that in QCD the inequalities imply chiral symmetry
breaking.

The presence of massless composite fermions in the supersymmetric case
has a natural explanation in terms of supersymmetry.
For  $\Nf \geq \Nc$, SQCD contains a manifold of degenerate vacuum
states.  The fluctuations along the flat directions of the potential
are described by effective scalar fields, and these scalar fields must
have supersymmetric partners, which are massless fermions.  Soft
supersymmetry breaking  removes the vacuum degeneracy and the flat
directions of the scalar potential.  Nevertheless, we saw that, in all
cases except for the baryon-number conserving vacuum of the $\notR$
case for $\Nf = \Nc$, the  massless composite fermions of the
supersymmetric limit remain massless after soft supersymmetry breaking.

 For the $N_f=N_c$ and $N_f=N_c+1$ cases, the massless fermions are
 gauge-invariant composite states.  They are required to remain
 massless in order to satisfy the 't Hooft anomaly matching conditions
 corresponding to unbroken chiral symmetries in the energetically
 preferred vacuum state.  This requirement is strong enough to keep
 the composite fermions massless even though their squark constituents
 obtain mass from soft supersymmetry breaking.

For $N_f > N_c+1$,  Seiberg argued that the $N=1$ SQCD theory  admits
a dual description in the infrared.  This dual theory contains a
dynamically generated gauge symmetry which is infrared-free for $\Nf \leq
{3\over 2} \Nc$ and possesses a
non-trivial infrared fixed point
for $3N_c > N_f > {3\over 2}N_c$.  The dual theory contains massless
composite fermions which belong to nontrivial representations of the
dual gauge symmetry.   We have argued that
 such a dual description
will  also exist  for the softly
broken theories, for some values of $N_f$ and
for small enough $m_Q$ and $m_g$.  The dual gauge theory is
infrared-free for $\Nf < {11\over 9} \Nc$ in the $R$ case and for
$\Nf < {9\over 7}\Nc$ in the $\notR$ case.  For
 ${11\over 9}N_c < N_f <
{11\over 2}N_c$ in the R case, and ${9\over 7}N_c < N_f < {9\over 2}N_c$
in the $\notR$ case, we expect to find a situation in which the theory
is controlled by a non-trivial infrared fixed point, with weak
coupling for the dual theory at the low-$\Nf$ boundary and weak
coupling for the original theory at the high-$\Nf$ boundary.
As in the supersymmetric case, the existence of this fixed point
can be proved near the boundary, that is, for large $\Nf$ and $\Nc$
approximately in the boundary ratios. It is likely that a single
infrared fixed point interpolates between these two boundary
situations.

The prospect of finding this kind of {\it infrared duality for
nonsupersymmetric
gauge theories}  is quite exciting. In the supersymmetric case we have
several arguments and cross-checks which support the
presence of the duality.   These include satisfaction of the
 't-Hooft anomaly matching conditions,
identification of all the gauge invariant operators in the chiral ring,
identification of all flat directions, and
verification of the behavior under mass perturbations \cite{NatiDual}.
So far the evidence for duality in the softly broken theories
relies only on the fact that the 't-Hooft anomaly matching conditions
are satisfied, and on their connection with the SQCD theory.
 For small supersymmetry breaking parameters,
the identification of those gauge invariant operators which were identified
in SQCD still goes through.
However, some operators which
were not identified in SQCD (such as the mesons made from the
dual quarks) apparently
should be identified after soft supersymmetry breaking. It seems that
the low energy spectrum after soft supersymmetry breaking
should remain the same as in SQCD, except for the splitting
between the states in a supermultiplet.  Hence, naively, we would expect
the operator identification to work in the same way.
Clearly, we would like to have more support for the nonsupersymmetric duality
conjecture. This is not easy in view of the fact that we have few tools
for analyzing
the non-perturbative behavior of the theory in the nonsupersymmetric
case.

(ii) {\it  In the decoupling limit most of  the
``exotic'' phenomena disappear.}

As we move towards the decoupling limit in which we take $m_Q$ (and
also $m_g$ in the $\notR$ case) to be large, it seems that most of  our
``exotic'' phenomena disappear. Typically, in these cases we encounter
a second order phase transition to the chirally broken phase of QCD.
This behavior is dictated by arguments that generalize mass inequalities
of vector-like gauge theories\cite{Weingarten,VafaWitten}.
For $\Nf < \Nc$, and for the baryon-number conserving vacuum in the
$\notR$ case for $\Nf = \Nc$, the decoupling limit to QCD is achieved
through a smooth transition from a
 softly broken vacuum which already exhibits the QCD
chiral symmetry breaking.  The corresponding $R$
case in this last model is ambiguous,
as described below.  In the other cases that we considered,
the decoupling limit is reached by a second-order phase transition
at some finite value of $m_Q$  in  which chiral symmetry is broken.
 Investigating
this phase transition is another interesting problem which we leave for
future research.

In the $R$ cases, it is possible that some exotic phenomena might
survive the decoupling limit.  For these theories, we presented two
options for the decoupling limit, option 2, with a conventional chiral
symmetry breaking pattern and no massless fermions, and option 1, with
the full chiral symmetry broken to $SU(\Nf)\times U(1)_B\times U(1)_R$
and a multiplet of massless fermions necessary to satisfy the 't Hooft
anomaly conditions for the unbroken $U(1)_R$.  The
 required composite fermions can be
constructed from massless quarks, antiquarks, and gauginos.  We have
not found any argument based on QCD inequalities to rule out this
possibility.  However, only in the
the baryon-number
conserving vacuum for $\Nf = \Nc$ in the  $R$ case did this
symmetry-breaking pattern arise naturally.  In all other cases, this
pattern still requires a second-order phase transition from the vacuum
which is preferred at small $m_Q$.

On top of the exotic behavior discussed above, there are further obvious
differences between the infrared
domain of the supersymmetric gauge theories and their decoupling
limits.
Here are several examples:
(1)   In the supersymmetric case the order parameters
associated with the
 chiral symmetry breaking  are expectation values of squark bilinear operators,
whereas in QCD  quark bilinears play this role. (2) Supersymmetric
fermionic baryons are composites of $N_c-1$ squarks and one quarks. (3)
Only totally anti-symmetric flavor representations  are relevant for the SQCD
baryons. In the cases in which the vacuum at small $m_Q$ can go
continuously into a vacuum of the decoupling limit,
we have found that  the order parameter is in fact a mixture of
the condensates of both bilinears, and
that it shows level-crossing behavior.
Close to the supersymmetric limit, the dominant component is the squark-squark
condensate.  As we go to the QCD limit, this
 contribution becomes negligible and
the quark-quark condensate takes over.  If option 2 for the $R$ case,
as described above, is realized, there is a related level-crossing
phenomenon,  in which  squark building blocks of composite fermions in
the supersymmetric limit are
replaced in the decoupling limit
  by a quark-gluino combination that has identical quantum numbers.

 (iii) {\it The exotic behavior
of the region  close to the supersymmetric limit should be detectable in
lattice simulations.}

 Simulations of softly broken SQCD should be
easier to perform than direct simulations of SQCD, since it is difficult to
maintain supersymmetry on the lattice. It may still be non-trivial  to
locate the region of the lattice coupling constants which reflects
softly broken SQCD, because this theory still has specific relations
among its renormalizable couplings.  However, we have argued that this
region can be found without unusual fine-tuning.  In particular,
we have discussed the issue of  possible
first order phase transitions in lattice gauge theories with scalars
 and indicated how
to avoid them.  With this barrier removed,
we expect the lattice simulations to reach the
continuum limit and reveal the rich structure of the exotic phenomena
described in this paper.

Finally, we list some additional issues which we have not resolved, and which
remain problems for future work:

The major  difficulty  encountered in passing from the
supersymmetric  gauge models
to QCD is the identification of
the SUSY
 breaking operators. It is usually
not easy to identify the relevant SUSY breaking operators in the
low energy effective potential description. As we explained in Section 2,
in softly broken supersymmetric theories we do have some control over this
problem. Following \cite{Evans}, we can show that our choice of the SUSY
breaking operators corresponds to those obtained from a supersymmetric theory
which includes some additional chiral superfields via spontaneous
supersymmetry breaking. In fact, with this approach one can relate the
resulting squark mass term to the K\"ahler kinetic term
in the underlying original SUSY theory. Thus, our lack of control over
the soft breaking terms is related to our lack of control over the
K\"ahler term. Clearly this question deserves further study.
We have also noted that other terms which may appear in the operator
identifications (such as e.g. $\tr \bigl[(T^\dagger
T)^n \bigr]$ for $n > 1$)
are typically suppressed by powers of $\Lambda$. Hence,
as long as we are considering vacua close to the origin, we are
justified in retaining only the lower terms we worked with. In these cases
we have more confidence in our results and can rely even on their
quantitative aspects. This is typically the situation for $N_f > N_c$.
However, when expectation values at the vacuum we are considering are
of order $\Lambda$ and higher, our neglect of the other higher terms
is not justified. This is the case for $N_f = N_c$, when the expectation
values are of order $\Lambda$, and for $N_f < N_c$ when the vacuum
of the theory runs to infinity in the supersymmetric limit.
 We believe that the
qualitative features of our results still hold in these cases,
but we certainly cannot
trust the quantitative aspects. This is the reason that in the $N_f=N_c$
case we could not decide which of the two possible vacuum states
 is preferred.

Another avenue of possible future research is the analysis of softly broken
supersymmetric gauge theories of other types, in particular, chiral
models, which also admit dual representations.  Recently, a number of
generalizations of Seiberg's original proposal have been
presented
\cite{Kutasov,NatiIntDual,youguys,KandS,IntPo,IntSPSO,Berk,LeStr,IntriLS}.
We expect the behavior of these theories upon adding soft supersymmetry
breaking terms to be similar to the behavior we found above for the
$SU(N_c)$ case.
Perhaps the study of these
theories will open even wider the unusual possibilities for
nonperturbative gauge dynamics.

  \Acknowledgements

We are grateful to Eliezer Rabinovici and the organizers of the 1995
Jerusalem Winter School for bringing us together and to Nathan Seiberg
for a stimulating set of lectures at the school which ignited our
interest in this
problem.  We thank Shmuel Nussinov and Gabrielle Veneziano for
stimulating conversations, and we are especially grateful to Tom Banks for
emphasizing the importance of QCD inequalities.  MEP thanks
the members of the high energy physics group at Tel Aviv
University for their hospitality  during the initial phase of this
work.


%
%

\appendix

\section{A QCD Inequality}

In this appendix we demonstrate an inequality which is useful in
understanding the limit of supersymmetric QCD in which the squark mass
is taken to infinity.  This limit is a vectorlike gauge theory of
quarks and the gluino, with no scalar fields.  We will show that, in
this limit, a flavor nonsinglet composite hadron
 cannot be massless if the pion is massive.  Our
argument is a straightforward  generalization of  arguments used to analyze
QCD by Weingarten \cite{Weingarten}.

To prove our claim, we follow Weingarten's proof that the baryon is heavier
than the pion.
Though Weingarten's original argument was given on the lattice (and
therefore was completely rigorous at the price of some complication),
we will apply a continuum version of the argument.  The crucial
observation is that, in vector-like gauge theories, the measure of
integration over gauge fields, which includes the determinants from the
integration over the fermions, is non-negative.  This can be seem
simply in the following way:
 For fermions of mass $m$, the fermion  determinant is
$\det(\Dslash_A+m)$ where
$\Dslash_A$ is the covariant derivative with gauge field $A$. In vector-like
theories this
is always positive, since the eigenvalues of $\Dslash_A$ are imaginary, and for
every
eigenvalue $ia$ with eigenvector $\psi_1$, $\gamma_5 \psi_1$ is an
eigenvector with
eigenvalue $-ia$, and the product of the contributions of both eigenvalues
to the
determinant is always positive. In the limit $m\to 0$ one could have
zero modes in gauge sectors of nontrivial Pontryagin number.  However,
these sectors do not contribute to any correlation function
we will consider.

In the analysis of Section 4.4, we are most concerned with the
possibility of a quark-antiquark-gluino bound state $\plp$, so let us
begin by considering this state.
Its propagator from $x$ to $y$ is given by
\beq
\int d\mu \sum_{a,a',b,b'} S_{\psi_{a,a'}}\delta^{ii'} \cdot
S_{{\bar{\psi}}_{b,b'}}\delta_{jj'} \cdot
S_{\lambda_{a,a',b,b'}}
\eeq{prop}
where $S_{\psi}$, $S_{\bar\psi}$ and $S_{\lambda}$ are the
quark, anti--quark and gluino propagators in the presence of fixed
background gauge
fields (we do not write the space-time indices explicitly),
and  $d\mu$ is the measure of integration over the gauge fields (including the
fermion determinants).
Since the integration measure is positive, this is smaller than
\beq
\int d\mu (\sum_{a,a'} |S_{\psi_{a,a'}}|^2)^{\half}
(\sum_{a,a'} |S_{{\bar{\psi}}_{a,a'}}|^2)^{\half}
(\sum_{a,a',b,b'} |S_{\lambda_{a,a',b,b'}}|^2)^{\half}.
\eeq{fbound}
Next,
we use the H\"older inequality, which says that for any positive measure,
\beq
|\int d\mu (f g)| \leq (\int d\mu |f|^2)^{\half} (\int d\mu
|g|^2)^{\half},
\eeq{holder}
to bound the propagator from above by
\beq
(\int d\mu \sum_{a,a'} |S_{\psi_{a,a'}}|^2 )^{\half}
	(\int d\mu (\sum_{a,a',b,b'} |S_{\lambda_{a,a',b,b'}}|^2)
		   (\sum_{a,a'} |S_{\psi_{a,a'}}|^2 ))^{\half}.
\eeq{sbound}

The next stage is to interpret each of the integrals in \leqn{sbound} as some
correlation function. The first integral is
proportional to  the propagator of the
pion, $\psi^i \bar\psi_j$.
In general, a propagator falls asymptotically as $e^{-m|x-y|}$,
 where $m$ is the lowest mass possible in the
intermediate state.  For the first integral $m$ is the pion
mass.  The second integral can, at worst, approach a constant
asymptotically.  Thus, the correlation function of $\plp$ is bounded
above by a constant times $\exp(-m_\pi |x-y|)$, where $m_\pi$ is the
pion mass.  Then the mass of the quark-antiquark-gluino bound state
must be greater than $m_\pi$.  This argument goes through in the same
way for any flavor-nonsinglet bound state, which necessarily contains at
least one quark and one antiquark or at least $N_c$ quarks.

Since the gluino-ball is a flavor singlet, there is no QCD inequality
relating its mass to that of flavor nonsinglet bound states.  This
leaves an ambiguity that we are  not able to resolve.  It is this
ambiguity that leads to the presence of option 1 (unbroken $U(1)_R$) in the
cases $\Nf \ge \Nc$.



\end{document}